{}
{}

\documentclass[11pt,a4paper]{article}

\newif\ifpublic\publictrue
\newif\iffancy\fancytrue

\usepackage[a4paper,text={160mm,247mm},centering]{geometry}
\usepackage{setspace}
\usepackage{amsmath,amssymb, amsfonts, geometry}
\makeatletter
\providecommand*{\shuffle}{%
  \mathbin{\mathpalette\shuffle@{}}%
}
\newcommand*{\shuffle@}[2]{%
  \sbox0{$#1\vcenter{}$}%
  \kern .15\ht0 
  \rlap{\vrule height .25\ht0 depth 0pt width 2.5\ht0}%
  \raise.1\ht0\hbox to 2.5\ht0{%
    \vrule height 1.75\ht0 depth -.1\ht0 width .17\ht0 %
    \hfill
    \vrule height 1.75\ht0 depth -.1\ht0 width .17\ht0 %
    \hfill
    \vrule height 1.75\ht0 depth -.1\ht0 width .17\ht0 %
  }%
  \kern .15\ht0 
}
\makeatother

\usepackage[pdfencoding=auto,bookmarks=true,hyperfigures=true]{hyperref}
\usepackage{graphicx}
\usepackage{float}
\usepackage[nosort]{cite}
\usepackage{lmodern}
\usepackage{amsbsy}
\usepackage[utf8]{inputenc}
\usepackage[font=small,labelfont=bf]{caption}
\usepackage{diagbox}
\usepackage{array}

\usepackage[usenames,dvipsnames]{xcolor}
\definecolor{dgreen}{rgb}{0,0.70,0.30}
\definecolor{gold}{rgb}{0.85,.66,0}
\definecolor{purple}{rgb}{1.0,0.3,0.6}

\usepackage{tikz}
\usetikzlibrary{calc} 
\usetikzlibrary{patterns} 
\usetikzlibrary{decorations.pathreplacing} 
\usetikzlibrary{decorations.markings} 
\usetikzlibrary{decorations.pathmorphing} 
\usetikzlibrary{positioning}

\iffancy

\def\showkeysrefformat#1{{\normalfont\tiny\ttfamily#1}}
\makeatletter
\def\SK@@ref#1>#2\SK@{%
 {\@inlabelfalse\leavevmode\vbox to\z@{%
 \vss\SK@refcolor\rlap{\vrule\raise .75em%
  \hbox{\showkeysrefformat{#2}}}}}}
\makeatother
\fi

\numberwithin{equation}{section}

\providecommand{\href}[2]{#2}

\makeatletter
\def\mr@ignsp#1 {\ifx\:#1\@empty\else #1\expandafter\mr@ignsp\fi}%
\newcommand{\multiref}[1]{\begingroup
\xdef\mr@no@sparg{\expandafter\mr@ignsp#1 \: }%
\def\mr@comma{}%
\@for\mr@refs:=\mr@no@sparg\do{\mr@comma\def\mr@comma{,}\ref{\mr@refs}}%
\endgroup}
\renewcommand{\eqref}[1]{(\multiref{#1})}
\makeatother

\makeatletter
\newcommand{\namedref}[2]{\hyperref[#2]{#1~\ref*{#2}}}
\newcommand{\secref}{\@ifstar{\namedref{Section}}{\namedref{section}}}
\newcommand{\subsecref}{\@ifstar{\namedref{Subsection}}{\namedref{subsection}}}
\newcommand{\appref}{\@ifstar{\namedref{Appendix}}{\namedref{appendix}}}
\newcommand{\tabref}{\@ifstar{\namedref{Table}}{\namedref{table}}}
\newcommand{\figref}{\@ifstar{\namedref{Figure}}{\namedref{figure}}}
\makeatother

\newcommand{\eqn}[1]{eq.~\eqref{#1}}
\newcommand{\Eqn}[1]{Equation~\eqref{#1}}
\newcommand{\eqns}[2]{eqs.~\eqref{#1} and~\eqref{#2}}

\newcommand{\rcite}[1]{ref.~\cite{#1}}

\providecommand{\hypersetup}[1]{}
\providecommand{\texorpdfstring}[2]{#1}

\hypersetup{plainpages=false}
\hypersetup{pdfpagemode=UseNone}
\hypersetup{bookmarksnumbered=true}
\hypersetup{pdfstartview=FitH}
\hypersetup{colorlinks=false}
\hypersetup{citebordercolor={.5 1 .5}}
\hypersetup{urlbordercolor={.5 1 1}}
\hypersetup{linkbordercolor={1 .7 .7}}

\makeatletter
\let\@keywords\@empty
\let\@subject\@empty
\providecommand{\keywords}[1]{\gdef\@keywords{#1}}
\providecommand{\subject}[1]{\gdef\@subject{#1}}
\def\thetitle{\@title}
\def\theauthor{\@author}
\def\thesubject{\@subject}
\def\thedate{\@date}
\def\thekeywords{\@keywords}
\makeatother
\AtBeginDocument{
\hypersetup{pdftitle={\thetitle}}%
\hypersetup{pdfauthor={\theauthor}}%
\hypersetup{pdfsubject={\thesubject}}%
\hypersetup{pdfkeywords={\thekeywords}}%
}

\newif\ifnote 
\notetrue

\allowdisplaybreaks

\newcommand{\ba}  {\begin{array}}
\newcommand{\ea}  {\end{array}}
\newcommand{\bdm} {\begin{displaymath}}
\newcommand{\edm} {\end{displaymath}}

\newcommand{\bea} {\begin{equation}\ba{lcl}}
\newcommand{\eea} {\ea\end{equation}}

\newcommand{\bc}  {\begin{center}}
\newcommand{\ec}  {\end{center}}

\def\beq{\begin{equation}}
\def\eeq{\end{equation}}

\def\Im{{\rm Im\,}}

\newcommand{\vecb}{\left(\begin{array}{c}}
\newcommand{\vece}{\end{array}\right)}
\newcommand{\ccb}{\left(\begin{array}{cc}}
\newcommand{\cce}{\end{array}\right)}
\newcommand{\cccb}{\left(\begin{array}{ccc}}
\newcommand{\ccce}{\end{array}\right)}
\newcommand{\ccccb}{\left(\begin{array}{cccc}}
\newcommand{\cccce}{\end{array}\right)}
\newcommand{\cccccb}{\left(\begin{array}{ccccc}}
\newcommand{\ccccce}{\end{array}\right)}

\newcommand{\pd}{\partial}

\newcommand{\al}{\alpha}

\newcommand{\z}{\zeta}

\newcommand{\te}{\textrm}

\newcommand{\dd}{\mathrm{d}}

\newcommand{\CF}{\mathcal F}
\DeclareMathOperator{\CI}{\mathcal I}

\newcommand{\CL}{\mathcal L}
\newcommand{\CN}{\mathcal N}
\newcommand{\CO}{\mathcal O}
\newcommand{\CP}{\mathcal P}

\def\MHV{\mathrm{MHV}}
\def\NMHV{\mathrm{NMHV}}

\def\reg{\textrm{reg}}
\def\MRK{\textrm{MRK}}

\newcommand{\nnl}{\nonumber\\}

\DeclareMathOperator{\zm}{\zeta}

\DeclareMathOperator{\omm}{\omega}

\DeclareMathOperator{\sgn}{\mathrm{sgn}}

\newcommand{\zfree}{{\zeta-\mathrm{free}}}
\newcommand{\xp}[1]{\chi_+^{#1}}
\newcommand{\xm}[1]{\chi_-^{#1}}
\newcommand{\Ep}{E_\psi}

\DeclareMathOperator{\Ib}{\mathcal{I}_\bullet}
\DeclareMathOperator{\Iz}{\mathcal{I}_0}
\DeclareMathOperator{\Ibz}{\mathcal{I}_{\bullet+0}}

\usepackage{amsthm} 
\theoremstyle{plain}

\vspace*{1.8cm}
\title{\textbf{Six-point remainder function in multi-Regge-kinematics: an efficient approach in momentum space}}
\author{
Johannes Broedel$^{\te{a,b}}$
, Martin Sprenger$^{\te{a}}$
}
\date{\today}

\begin{document}
\pdfbookmark[1]{Title Page}{title} \thispagestyle{empty}
\begin{flushright}
  \verb!HU-EP-15/61!\\
  \verb!HU-Mathematik-2015-15!
\end{flushright}
\vspace*{0.4cm}

\begin{center}%
  \begingroup\LARGE\bfseries\thetitle\par\endgroup
  \vspace{1.4cm}

\begingroup\large\theauthor\par\endgroup
\vspace{8mm}%

\begingroup\itshape
$^{\te{a}}$Institut f\"ur Theoretische Physik,\\
Eidgen\"ossische Technische Hochschule Z\"urich\\
Wolfgang-Pauli-Strasse 27, 8093 Z\"urich, Switzerland
\par\endgroup
\vspace{4mm}
\begingroup\itshape
$^{\te{b}}$Institut f\"ur Mathematik und Institut f\"ur Physik,\\
Humboldt-Universit\"at zu Berlin\\
IRIS Adlershof, Zum Gro\ss{}en Windkanal 6, 12489 Berlin, Germany
\par\endgroup
\vspace{4mm}

\vspace{0.8cm}

\begingroup\ttfamily
jbroedel@physik.hu-berlin.de, 
sprengerm@itp.phys.ethz.ch
\par\endgroup

\vspace{1.2cm}

\bigskip

\textbf{Abstract}\vspace{5mm}

\begin{minipage}{13.4cm}
Starting from the known all-order expressions for the BFKL eigenvalue and
impact factor, we establish a formalism allowing the direct calculation of the
six-point remainder function in $\CN=4$ super-Yang--Mills theory in momentum
space to -- in principle -- all orders in perturbation theory. Based upon
identities which relate different integrals contributing to the inverse
Fourier-Mellin transform recursively, the formalism allows to easily access the
full remainder function in multi-Regge kinematics up to $7$ loops and up to
$10$ loops in the fourth logarithmic order. Using the formalism, we
prove the all-loop formula for the leading logarithmic approximation proposed
by Pennington and investigate the behavior of several newly calculated
functions. 
\end{minipage}

\vspace*{4cm}

\end{center}

\newpage

\setcounter{tocdepth}{2}
\tableofcontents


\section{Introduction}
\label{sec:introduction}

Starting with the seminal papers of Balitsky, Fadin, Kuraev and Lipatov (BFKL)
\cite{Lipatov:1976zz, Fadin:1975cb, Kuraev:1976ge, Balitsky:1978ic}, the study
of the multi-Regge regime of scattering amplitudes in non-Abelian gauge
theories has a long and rich history.  Besides being of phenomenological
importance, it is in this kinematical regime that integrable structures and
relations to spin chain models were identified first \cite{Lipatov:1993yb,
Lipatov:1994xy, Faddeev:1994zg}.  A key feature of the multi-Regge regime is
that the perturbative expansion is reorganized naturally: instead of
expanding in the number of loops, amplitudes in the multi-Regge limit are
usually expanded in logarithmic orders, starting with the leading logarithmic
approximation (or LLA for short) and followed by N$^k$LLA \mbox{(for
(Next-to)$^k$-LLA)} which are subleading in kinematics.  Hence, each
logarithmic order contains information from all loop orders and therefore
from finite coupling, albeit in a very specific way.  

Multi-Regge kinematics (MRK) is thus a natural regime to study in planar
$\CN=4$ super Yang-Mills theory, which is believed to be solvable at finite
coupling, as well. Indeed, the study of the multi-Regge regime in this theory
revealed the failure of the Bern--Dixon--Smirnov (BDS) ansatz
\cite{Bern:2005iz} for the six-gluon MHV amplitude in
refs.~\cite{Bartels:2008sc, Bartels:2008ce} before the explicit two-loop
calculation of the remainder function appeared in refs.~\cite{DelDuca:2009au,
DelDuca:2010zg}.  Further results include the calculation of the remainder
function at strong coupling \cite{Bartels:2010ej, Bartels:2012gq,
Bartels:2013dja, Bartels:2014ppa, Bartels:2014mka}, the study of higher-point
amplitudes \cite{Bartels:2011ge, Prygarin:2011gd, Bartels:2013jna,
Bartels:2014jya} and the identification of the correct basis functions
describing the multi-Regge limit \cite{Dixon:2012yy, Pennington:2012zj,
DelDuca:2013lma}.  Those functions will play a key role later on.

Direct calculations allowed the determination of the six-gluon MHV and NMHV
remainder function in Fourier-Mellin space up to NLLA \cite{Lipatov:2010ad,
Fadin:2011we, Lipatov:2012gk}.  However, modern amplitude techniques such as
the amplitude bootstrap in \rcite{Dixon:2014voa} allowed to push the known
results up to N$^2$LLA (and N$^3$LLA for the impact factor). The efforts to
determine the remainder function to increasing accuracy culminated in a recent
paper \cite{Basso:2014pla} in which conjectures for exact expressions in the
multi-Regge regime are put forward.  These expressions are derived by an
analytic continuation from the collinear regime which is described by the
Wilson loop OPE \cite{Basso:2013vsa, Basso:2013aha, Basso:2014koa,
Basso:2014nra} and reproduce the known data at weak and strong coupling.

These equations and their solutions are, however, formulated in Fourier-Mellin space.
The calculation of the remainder function in momentum space therefore still
requires carrying out an --  in general -- tedious inverse Fourier-Mellin transform.
The problem is simplified by the observation made in \cite{Dixon:2012yy} that
all integrals appearing yield a special class of functions, namely the
single-valued harmonic polylogarithms (SVHPLs) constructed in ref.~\cite{Brown2004527}.
This, together with symmetry considerations, allows to write down a compact
ansatz for the result of each integral, which can then be matched with the
first orders of the integral's series expansion. While feasible for low
weights, the matching becomes cumbersome for large weights. 

Another step forward was made in \rcite{Drummond:2015jea}: there it was shown
that the analytic continuation from the collinear regime to the multi-Regge
regime can be carried out term-by-term.  More importantly, the authors use an
observation made in \rcite{Drummond:2013nda} to argue that in carrying out the
inverse Fourier-Mellin transform it is sufficient to focus on a specific subset
of terms. Demanding single-valuedness then restores the full answer.
This method, too, is algorithmic in nature, but is not formulated in momentum space.
The procedure is explained in \secref{sec:basics} below.

In this paper, we combine the implications of single-valuedness alluded to
above with relations between different integrals: considering the residues from
which the single-valued result can be obtained, we can trace back all integrals to
a small set of trivial basis integrals. Based on those relations, we propose an
algorithm for the calculation of the six-point remainder function at any loop
order and any logarithmic order. The resulting expressions are generated
directly in momentum space by operations acting on SVHPLs naturally, namely by
appending and prepending indices and applying shuffles. In this sense, this
paper is a natural extension of the work \cite{Pennington:2012zj}, in which
all-loop formul\ae{} for LLA in momentum space were conjectured, to all
logarithmic orders.  In fact, we are able to proof Pennington's formula for LLA
based on our algorithm.

This paper is organized as follows.  In \secref{sec:basics} we review the
relevant formul\ae{} to introduce notation and describe some key features of
SVHPLs. \secref*{sec:solution} contains an elaborate description of our
formalism in the MHV case along with several examples, and is followed by a
discussion of the NMHV case in \secref{sec:nmhv}.  Based on our
algorithm, we extend the known results for N$^3$LLA and N$^4$LLA up to $10$ and
$9$ loops, respectively, in \secref{sec:app_hl} and consider the
collinear-Regge limit as an application of our results in
\secref{sec:app_crl}.  Finally, we provide a proof of the conjectured all-loop LLA
formul\ae{} put forward in \rcite{Pennington:2012zj} in \secref{sec:proof} before
summarizing our results in \secref{sec:conclusion}.


\section{BFKL/Remainder function basics}
\label{sec:basics}
We are interested in the six-point remainder functions $R_6^\MHV$ and
$R_6^\NMHV$ of $\mathcal{N}=4$ super-Yang--Mills theory, which describe the
deviation of the full amplitude from the BDS ansatz \cite{Bern:2005iz}, 
\begin{subequations}
\begin{align}
  A^\MHV_6&=A_{\text{BDS}}\,e^{R_6^\MHV}\label{eqn:BDSR}\,,\\
  A^\NMHV_6&=A_{\text{BDS}}\,R_6^\NMHV\label{eqn:BDSRNMHV}.
\end{align}
\end{subequations}
Note that in defining the remainder
function for NMHV we follow the conventions in \rcite{Dixon:2014iba}.  Switching
from the external momenta $k_i$ to dual variables,
$k_i=:x_i-x_{i+1}=:x_{ii+1}$, the multi-Regge limit of the six-gluon remainder
functions is described by the following behavior of the dual conformal cross
ratios:
\begin{equation}
	u_1:=\frac{x_{13}^2x_{46}^2}{x_{14}^2x_{36}^2}\rightarrow 1, \quad u_2:=\frac{x_{24}^2x_{15}^2}{x_{25}^2x_{14}^2}\rightarrow 0, \quad u_3:= \frac{x_{35}^2x_{26}^2}{x_{36}^2x_{25}^2}\rightarrow 0,
	\label{eq:mrl_crs}
\end{equation}
such that the reduced cross ratios
\begin{equation}
	\tilde{u}_2:=\frac{u_2}{1-u_1}=:\frac{1}{|1+w|^2},\quad \tilde{u}_3:=\frac{u_3}{1-u_1}=:\frac{|w|^2}{|1+w|^2}
	\label{eq:mrl_red_crs}
\end{equation}
remain finite. As usual, we describe the kinematics with the large cross ratio
$u_1$ and the parameters $w$, $w^*$, which parameterize the remaining
kinematical freedom in the multi-Regge regime.  Taking the limit
\eqn{eq:mrl_crs} of the cross ratios na\"ively will lead to a vanishing
remainder function.  However, one obtains a non-trivial result by first
performing the analytic continuation
\begin{equation}
	u_1\rightarrow e^{-2i\pi}u_1
\end{equation}
to the so-called Mandelstam region before taking the limit (\ref{eq:mrl_crs}).

While the kinematical setup is universal for the six-point remainder functions,
the integrals appearing in the calculation depend on the helicity
configuration. Therefore the discussion in the current as well in the next
section will be conducted on the example of $R_6^\MHV$, while the appropriate
modifications in order to obtain $R_6^\NMHV$ are described in
\secref{sec:nmhv}.

In the Mandelstam region the six-point MHV\footnote{Note that in all-outgoing
convention this corresponds to the helicity configuration $(++-++-)$, where
legs $1$,$2$,$3$ and $6$ are the high-energy lines.} remainder function in
multi-Regge kinematics can be determined from the following integral
\cite{Bartels:2008sc, Lipatov:2010ad, Fadin:2011we}:
\begin{align}
  e^{R_6^\MHV+i\pi\delta}|_\MRK=&\cos \pi\omm_{ab}+i\frac{a}{2}\sum_{n=-\infty}^{\infty}(-1)^n
  \left(\frac{w}{w^*}\right)^{\frac{n}{2}}\int_{-\infty}^\infty\frac{\dd\nu}{\nu^2+\frac{n^2}{4}}\,|w|^{2i\nu}\Phi_\reg^\MHV(\nu,n)\nnl
  &\qquad\qquad\qquad\times\,\exp\left[-\omm(\nu,n)\left(\log(1-u_1)+i\pi+\frac{1}{2}\log\frac{|w|^2}{|1+w|^4}\right)\right]\,.
  \label{eqn:master}
\end{align}
The two contributions in \eqn{eqn:master} have a nice physical
interpretation: the first term describes the contribution of a Regge pole
exchange while the second term describes the contribution of a Regge cut
generated by a bound state of two Reggeons.  In the latter contribution the
only unknown quantities are the impact factor $\Phi^\MHV_\reg(\nu,n)$ describing the
coupling of the bound state of Reggeons to the produced gluons and the BFKL
eigenvalue $\omm(\nu,n)$ describing the evolution of the bound state.  Both
quantities have a loop expansion in powers of $a$:
\begin{align}
  \omm(\nu,n)&=-a(E^{(0)}+aE^{(1)}+a^2E^{(2)}+\CO(a^3)),\nnl
  \Phi_\reg^\MHV&=1+a\,\Phi^{(1),\MHV}+a^2\,\Phi^{(2),\MHV}+\CO(a^3)\,.
  \label{eqn:BFKLloop}
\end{align}
Furthermore, $\delta$ corresponds to a phase which in the full amplitude
\eqn{eqn:BDSR} cancels with a similar contribution arising from the BDS ansatz
\cite{Bern:2005iz}.  Both $\delta$ and the Regge pole contribution
$\omm_{ab}$ are related to the cusp anomalous dimension $\gamma_K$ via
\begin{align}
  \omm_{ab}=\frac{1}{8}\gamma_K(a)\log|w|^2\quad \text{and} \quad \delta=\frac{1}{8}\gamma_K(a)\log\frac{|w|^2}{|1+w|^4}
\end{align}
and are therefore known to all orders in perturbation theory
\cite{Beisert:2006ez}.  

Due to the behavior of the cross ratio $u_1$ in the multi-Regge limit
\eqn{eq:mrl_crs}, the expression $\log(1-u_1)$ in the integrand of
\eqn{eqn:master} is large and the remainder function can be conveniently
organized in powers of this logarithm at each loop order.  In
particular, 
\begin{equation}
  R_6^\MHV\big|_\MRK=2\pi i\sum\limits_{\ell=2}^\infty\sum\limits_{n=0}^{\ell-1}a^\ell \log^n(1-u_1)\Big[g^{(\ell)}_n(w,w^*) + 2 \pi i \,h^{(\ell)}_n(w,w^*)\Big]\,.
  \label{eq:rem_exp_gh}
\end{equation}
In \eqn{eq:rem_exp_gh}, all terms with $n=\ell-1$ are referred to as the
leading logarithmic approximation (LLA) and similarly the terms with
$n=\ell-1-k$ are called (Next-to)$^k$-LLA.  Fortunately, the imaginary and real
parts $g^{(\ell)}_n$ and $h^{(\ell)}_n$ are not independent
\cite{Dixon:2012yy}: all functions $h^{(\ell)}_n$ can be determined from the
imaginary parts $g^{(\ell)}_n$. Thus the knowledge of all functions
$g^{(\ell)}_n$ will pin down the remainder function.

Accordingly, the problem of calculating the remainder function $R^\MHV_6|_\MRK$
via \eqn{eqn:master} boils down to evaluating the real part of the sum over the
integral. The crucial ingredients here are the loop expansions of the impact
factor $\Phi^\MHV_\reg(\nu,n)$ and the BFKL eigenvalue $\omm(\nu,n)$ in
\eqn{eqn:BFKLloop}.  The first two orders of these quantities are known from
direct calculation \cite{Bartels:2008sc, Lipatov:2010ad, Fadin:2011we}, while
one further order of the BFKL eigenvalue and two further orders for the impact
factor could be determined from the amplitude bootstrap in
refs.~\cite{Dixon:2012yy, Dixon:2014voa}.  A general solution, however, was
found by Basso, Caron-Huot and Sever \cite{Basso:2014pla} only recently: the
key idea is to link the solution of \eqn{eqn:BFKLloop} to quantities appearing
in the Wilson loop OPE \cite{Basso:2013vsa, Basso:2013aha, Basso:2014koa,
Basso:2014nra}, which are known exactly from integrability, via analytic
continuation. Using this method, the eigenvalue $\omm$ and impact factor
$\Phi^\MHV_\reg$ can be calculated using the solution of the BES equation
\cite{Beisert:2006ez} to arbitrary order in the coupling constant $a$ in
principle. 

Higher orders of the BFKL eigenvalue and the impact factor are usually expressed
in terms of the quantities $N$ and $V$ \cite{Dixon:2012yy}
\begin{equation}
	N:=\frac{n}{\nu^2+\frac{|n|^2}{4}}\quad\text{and}\quad V:=\frac{i\nu}{\nu^2+\frac{|n|^2}{4}}\,,
  \label{eqn:definitions}
\end{equation}
the derivative
\begin{equation}
  D_\nu:=-i\pd_\nu\,
  \label{eqn:derivative}
\end{equation}
and the lowest order of the BFKL eigenvalue 
\begin{align}
  E^{(0)}&=-\frac{1}{2}\sgn(n)N+\psi\left(1+i\nu+\frac{|n|}{2}\right)+\psi\left(1-i\nu+\frac{|n|}{2}\right)-2\psi(1)\,.
  \label{eqn:E0}
\end{align}
In this language, the first orders of eigenvalue (beyond \eqn{eqn:E0}) and
impact factor take the form 
\begin{align}
  E^{(1)}&=-\frac{1}{4}D_\nu^2 E^{(0)}+\frac{1}{2}V D_\nu E^{(0)}-\zm_2E^{(0)}-3\zm_3,\nnl
  E^{(2)}&=\frac{1}{8}\Big(\frac{1}{6}D_\nu^4E^{(0)}-VD_\nu^3E^{(0)}+\big(V^2+2\zm_2\big)D_\nu^2E^{(0)}-V(N^2+8\zm_2)D_\nu E^{(0)}\nnl
		&\qquad\qquad\qquad+\zm_3(4V^2+N^2)+44\zm_4E^{(0)}+16\zm_2\zm_3+80\zm_5 \Big),\nnl
  \Phi^{(1),\MHV}&=-\frac{1}{2}\big(E^{(0)}\big)^2 - \frac{3}{8} N^2 - \zm_2,\nnl
  \Phi^{(2),\MHV}&=\frac{1}{2} \big(\Phi^{(1),\MHV}\big)^2 - E^{(1)}E^{(0)} + \frac{1}{8} \big(D_\nu E^{(0)}\big)^2 + \frac{5}{64}N^2 (N^2 + 4 V^2)\nnl
	 &\qquad\qquad\qquad-\frac{\zm_2}{4} \left(2 (E^{(0)})^2 + N^2 + 6 V^2\right) + \frac{17\zm_4}{4}.
\end{align}
For reasons to become clear below we will later replace the quantities $N$ and
$V$ with new variables $\xp{}$ and $\xm{}$:
\begin{equation}
  \xp{}:=\left(V+\frac{N}{2}\right)\quad\text{and}\quad \xm{}:=\left(V-\frac{N}{2}\right)\,.
  \label{eqn:xpxm}
\end{equation}
As discussed in \rcite{Dixon:2012yy}, the functions $g_n^{(\ell)}$ can be
expressed in terms of single-valued harmonic polylogarithms (SVHPLs)
\cite{Brown2004527}. SVHPLs are linear combinations of products of usual
harmonic polylogarithms (HPLs) in the variables $z$ and $z^*$, which are
combined in such a way that their branch cuts cancel.  Different
SVHPLs\footnote{As will be clear from the examples given in the text, the
$\CL_s(z)$ are functions of both $z$ and $z^*$. It is, however, customary, to
simply denote the functions $\CL_s(z)$.} $\CL_s(z)=\CL_s$  are labeled by a
word $s$ built from the alphabet $\lbrace x_0,x_1 \rbrace$, which is reflected
in the identification
\begin{equation}
        x_{i_1}\dots x_{i_n}\leftrightarrow\mathcal{L}_{i_1\dots i_n}(z).
	\label{eq:rep_xl}
\end{equation}
Whenever necessary, a word built from letters $x_0$ and $x_1$ is understood to be
replaced by the appropriate polylogarithm using the above assignment.

As derived by Brown in \rcite{Brown2004527}, the particular combinations of
HPLs resulting in SVHPLs are unique and can be extracted from demanding
triviality of the monodromies around singular points of HPLs.  Nicely, this
single-valuedness condition can be expressed as an equation involving the
Drinfeld associator.  Explicitly, the SVHPLs with lowest weights are given by
\begin{align}
  \CL_0(z)&=H_0(z)+H_0(z^*)\nnl
  \CL_1(z)&=H_1(z)+H_1(z^*)\nnl
  \CL_{00}(z)&=H_{00}(z)+H_{00}(z^*)+H_0(z)H_0(z^*)\nnl
  \CL_{10}(z)&=H_{10}(z)+H_{01}(z^*)+H_1(z)H_0(z^*)\nnl
  \CL_{101}(z)&=H_{101}(z)+H_{101}(z^*)+H_{10}(z)H_1(z^*)+H_{1}(z)H_{10}(z^*)\nnl
  &\quad\vdots
  \label{eqn:svhpl}
\end{align}
With increasing weight, these expressions become more complex and
$\zm$-values enter explicitly, for example
\begin{align}
  \CL_{1010}(z)&=H_{1010}(z)+H_{0101}(z^*)+H_{101}(z)H_{0}(z^*)+H_{1}(z)H_{010}(z^*)\nnl
	       &\qquad+H_{10}(z)H_{01}(z^*)-4\zm_3H_1(z^*).
	       \label{eq:ex_zeta_svhpl}
\end{align}
An elaborate introduction to SVHPLs in the context of the six-point
remainder function in MRK can be found in section 3 of \rcite{Dixon:2012yy}.
Here, we will simply describe the essential properties of SVHPLs for the sections to follow.
First, two SVHPLs labeled by words $s_1$ and $s_2$ satisfy the shuffle relation
\begin{equation}
  \CL_{s_1}(z)\,\CL_{s_2}(z)=\sum\limits_{s\in s_1\shuffle s_2}\CL_s(z)\,,
\end{equation}
where $s_1\shuffle s_2$ denotes all permutations of $s_1 \cup s_2$ which
preserve the order of elements in $s_1$ and $s_2$.  Furthermore, the SVHPLs
satisfy differential equations which can be most easily stated by introducing a
generating functional for the SVHPLs,
\begin{equation}
  \CL(z)=\sum\limits_{s\in X^*}\CL_s(z)s=1+\CL_0(z)\,x_0+\CL_1(z)\,x_1+\CL_{00}(z)\,x_0x_0+\CL_{01}(z)\,x_0x_1+\dots\,,
	\label{eq:genfunc}
\end{equation}
where $X^*$ are all words in the alphabet $\lbrace x_0,x_1\rbrace$.  This
generating functional satisfies the differential equations
\begin{equation}
  \frac{\pd}{\pd z}\CL(z)=\left(\frac{x_0}{z}+\frac{x_1}{1-z}\right)\CL(z),\quad 
  \frac{\pd}{\pd z^*}\CL(z)=\CL(z)\left(\frac{y_0}{z^*}+\frac{y_1}{1-z^*}\right)\,,
	\label{eqn:svhpldiff}
\end{equation}
where $\lbrace y_0,y_1\rbrace$ is an additional alphabet, which appears in the
construction of ref.\cite{Brown2004527} and is related by the single-valuedness
condition to the alphabet $\lbrace x_0,x_1\rbrace$ introduced above. Solving
this condition order by order, one finds
\begin{align}
  y_0&=x_0\quad\text{and}\nnl
  y_1&=x_1-\z_{3}(2x_0x_0x_1x_1-4x_0x_1x_0x_1+2x_0x_1x_1x_1+4x_1x_0x_1x_0+\cdots)+\cdots\,,
  \label{eqn:fixpointsol}
\end{align}
where the words appearing in the deviation of $y_1$ from $x_1$ always include
at least two $x_1$'s and are always multiplied by a $\zm$-value.  From
\eqn{eqn:svhpldiff} we see that derivatives with respect to $z$ act by taking
off the first index of a given SVHPL and multiplication with some factor.
However, due to the non-trivial relation between the two alphabets, taking
derivatives with respect to $z^*$ does not simply act by chopping off the last
index, but can lead to additional terms which always include $\zm$-values.
These additional terms constitute one of the main challenges of this paper and
will be discussed in detail later on.

When necessary, we will employ the collapsed notation for HPLs and SVHPLs,
which is related to the notation with explicit indices via
\begin{equation}
	\CL_{a_1,a_2,\dots}\leftrightarrow\CL_{\underbrace{\scriptstyle 0\dots0}_{a_1-1}1\underbrace{\scriptstyle 0\dots0}_{a_2-1}1\dots}
	\label{eq:def_collapsed}
\end{equation}
and equivalently for the HPLs.  Note that in the collapsed notation commas are
used to separate the indices, while in the explicit index notation no commas
are present.\par\medskip

By either direct evaluation (for low orders) or by matching a suitable ansatz
with the analytic properties and the class of integrals available, the
functions $g^{(\ell)}_n$ have been determined completely up to five loops
\cite{Dixon:2012yy, Dixon:2014voa, Drummond:2015jea} while results for LLA and
NLLA are available for even higher loops.  Note that direct evaluation of
\eqn{eqn:master} amounts to closing the integration contour at infinity in the
positive or negative half-plane, depending on whether $|w|\lessgtr 1$, and
summing up infinitely many residues, which for a given $n$ are located at
$i\nu=\pm(|n|/2+m)$, where $m$ is a positive integer.

A substantial improvement compared to earlier calculations made the
determination of \eqn{eqn:master} possible for N${}^3$LLA and N${}^4$LLA at the
five-loop level: based on an observation made in \rcite{Drummond:2013nda}, the
authors of \rcite{Drummond:2015jea} recognized that the result in MRK can be
obtained (up to subtleties to be discussed below) by calculating the result in
a yet different kinematical limit on top of the MRK: the so-called
double-scaling limit, $w^*\rightarrow 0$ with $w$ fixed.
Once promoting the resulting harmonic polylogarithms to SVHPLs by simply
replacing  $H\rightarrow\CL$, the six-point remainder function in MRK is nicely
recovered from the double-scaling limit.  The mechanism works upon assuming
single-valuedness of the result and taking into account that the leading
contribution to a SVHPL in the limit $w^*\to 0$, once we simply drop all
logarithms $\log w^*$, is the corresponding harmonic polylogarithm
(cf.~\eqn{eqn:svhpl}) with the same index structure. 

Comparing with \eqn{eqn:master}, we see that we can implement this prescription
by setting $w^*=1$ in the integrand, as well as only calculating the residue at
$i\nu=\pm n/2$.  This automatically drops all logarithms $\log w^*$ and by
focusing on the residues at $i\nu=\pm n/2$ we extract the $w^*$-independent
part of the integral, which is the leading term in the limit $w^*\rightarrow
0$.  The $w^*$-dependence is stored in the residues at $i\nu=\pm(n/2+m)$, where
$m$ is a positive integer, which we neglect.  In this way, the integral will
evaluate to HPLs, which we can then promote to their single-valued analogues to
obtain the full result.  In the following section we will thoroughly describe
our method. 


\section{Solving the integral}
\label{sec:solution}

Having the different approaches described above for evaluating \eqn{eqn:master}
at our disposal, it seems rather straightforward to evaluate the remainder
function to high logarithmic and loop order. However, already at five loops the
algebraic effort in matching a suitable ansatz to the sums resulting from the
evaluation of the integral is considerable. Loop-orders higher than five are
possible for low logarithmic orders, because those come with a rather small
number of terms in the integrand. A solution for six loops including all
logarithmic orders, however, seems out of reach. Removing this obstruction is
precisely the main point of this article. There are two concepts our formalism
is based upon, which will be described immediately.

\paragraph{Residues.} The first ingredient is based on the observation made in
refs.~\cite{Drummond:2015jea,Drummond:2013nda} and described in the previous
subsection: the full result of \eqn{eqn:master} can be obtained by closing the
integration contour \textit{below}\footnote{Closing the contour below corresponds to
the case where $|w|<1$. We could equally well consider the case $|w|>1$ and
close the contour above, which would lead to the same results.} the real axis
and considering the residue at $i\nu=n/2$ with $n\geq 0$. In
contradistinction to \rcite{Drummond:2015jea}, where the calculation is
performed in the double-scaling limit, we are working directly in MRK:
therefore we need to consider half of the residue at $\nu=n=0$ in addition \cite{Bartels:2008ce}.
As alluded to above, in \rcite{Drummond:2015jea}, this contribution gets automatically
restored by extending the result from the double-scaling limit to MRK. 
Following \rcite{Dixon:2012yy}, we define a shorthand for the integrals in
question:
\begin{equation}
  \CI[\CF(\nu,n)]=\frac{1}{\pi}\sum\limits_{n=-\infty}^{\infty}(-1)^n\left(\frac{w}{w^*}\right)^{\frac{n}{2}}\int_{-\infty}^\infty\frac{\dd\nu}{\nu^2+\frac{n^2}{4}}\,|w|^{2i\nu}\,\CF(\nu,n) \,.
  \label{eqn:defI}
\end{equation}
In order to further simplify the notation below we introduce:
\begin{equation}
  \Ib[\CF(\nu,n)]=\CI[\CF(\nu,n)]\big|_{i\nu=\frac{n}{2},w^*=1}\quad\text{and}\quad 
  \Iz[\CF(\nu,n)]=\frac{1}{2}\CI[\CF(\nu,n)]\big|_{\nu=n=0,w^*=1},
  \label{eqn:defIbIz}
\end{equation}
where the above notation shall imply that in evaluating the integral only the
contribution from residues noted are taken into account. Furthermore, the
prescription in which only half of the residue at $\nu=n=0$ contributes to the
integral is already included in the above definition. Naturally, the sum over
$n$ will collapse in this case. Employing the above notation, the general
technique used in this paper can be abbreviated by
\begin{equation}
  \CI[\CF(\nu,n)]=
  \left(\Ib[\CF(\nu,n)]+\Iz[\CF(\nu,n)]\right)\Big|_{H\to\CL}\,.
  \label{eq:res_presc}
\end{equation}
Considering only poles in the lower half-plane or at $\nu=n=0$ is obviously
simpler compared to the evaluation of the full integral: in particular we get
rid of the dependence on the signature of $n$, as all $n$'s in question are
positive. Further simplifications in the classification of integrals below will
occur upon splitting the lowest-order BFKL eigenvalue 
\begin{equation}
	E^{(0)}=-\frac{1}{2}(\xp{}-\xm{})+\Ep
	\label{eq:e0_split}
\end{equation}
into a part originating from the first term in \eqn{eqn:E0} and into a part
$\Ep$ containing the polygamma functions: while the former one does raise the
order of the pole at $i\nu=n/2$ in \eqn{eqn:master}, the contribution $\Ep$ is
free of poles at $i\nu=n/2$.  Note again that \eqn{eq:e0_split} only holds
because we are considering poles with $n\geq 0$ exclusively, as mentioned
above.

\paragraph{Relations between residues.} The second important ingredient to our
formalism is to exploit relations between residues.  In particular, we will
derive a couple of rules which relate integrals among each other.  In
evaluating their residues, it will be very useful to rewrite particular values
of (derivatives of) polygamma functions in terms of Euler $Z$-sums. Relating
residues is the key to the efficiency of our method: instead of calculating
numerous integrals by determining fudge coefficients of an ansatz, we calculate
only very few integrals and express the residues of all other integrals in
terms of those. The relations between the residues can be cast into a set of
simple rules to be discussed below.

The formalism for the evaluation of \eqn{eqn:master} can be summarized as follows:
\textit{
\begin{itemize}
   \setlength{\itemsep}{-2pt}
   \item we expand the integrand in \eqn{eqn:master} to the desired order in the
     coupling constant $a$ and use relation (\ref{eq:e0_split}) to obtain
     several integrals of the general form 
    \begin{equation}
      \CI\left[\xp{n_+}\xm{n_-}\left(\prod_i D_\nu^{a_i}\Ep\right)\Ep^k\right]\,.
      \label{eqn:integraltype}
    \end{equation}
  \item we relate residues at $i\nu=n/2$ of integrals of type
    (\ref{eqn:integraltype}) to residues of the basis
    integrals
    \begin{equation}
      \CI\left[\left(\prod_i D_\nu^{a_i}\Ep\right)\Ep^k\right]\,.
      \label{eqn:basisinte}
    \end{equation}
  \item for the residues at $\nu=n=0$ we use a similar set of rules to 
    again relate those to the residues at $\nu=n=0$ of the same basis
    integrals. 
  \item once all contributions to the integrals are calculated, we promote the harmonic
    polylogarithms to their single-valued cousins. 
\end{itemize}
}
Using this approach and calculating a couple of basis integrals upfront
allows a very fast determination of high orders of the remainder function. The
results thus obtained match all known expressions.  


\subsection{Basic mechanism}

Let us get started by evaluating the simplest integral of type
(\ref{eqn:integraltype}): $\CI[1]$. One easily finds
\begin{align}
  \Ib[1]=\sum_{n=1}^\infty\frac{2 (-w)^n}{n}=2H_1(-w)\,.
  \label{eqn:I1}
\end{align}
Similarly,
\begin{align}
  \Iz[1]=\log w=H_0(w)\,,
  \label{eqn:I1zero}
\end{align}
which leads -- after promoting the usual harmonic polylogarithms to their
single-valued analogues\footnote{Note that the sign difference in the arguments
of \eqns{eqn:I1}{eqn:I1zero} is not a problem - in promoting
$H_0\rightarrow \CL_0$ we get another term $\log w^*$ which combines with the
result of \eqn{eqn:I1zero} to $\log |w|^2$ for which the sign does not
matter.} -- to the known result 
\begin{equation}
	\CI[1]=\CL_0(-w)+2\CL_1(-w),
	\label{eq:res_I1}
\end{equation}
cf.~\eqn{eqn:svhpl} with the identification $(z,z^*)=(-w,-w^*)$.

\subsection{Integrals of the form $\CI[\xp{n_+}\xm{n_-}]$}
\label{ssec:Ixpxm}
Integrals of the form $\CI[\xp{n_+}\xm{n_-}]$ are determined in such a way as
to ensure single-valuedness of the resulting polylogarithm $\CL_s(-w)$.
Comparing \eqn{eqn:svhpldiff} with the following relation for the
$w$-dependence of the integral $\CI$ defined in \eqn{eqn:defI}
\begin{equation}
	-w\frac{\pd}{\pd w}\left(\xm{}w^{i\nu+\frac{n}{2}}\right)=w^{i\nu+\frac{n}{2}}
\label{eqn:wdw}
\end{equation}
it is natural to expect an insertion of $\xm{}$ in the integrand to be
related to prepending $(-x_0)$ to the result of the integral
without the insertion. Accordingly, taking the derivative with respect to
$w^*$, one finds
\begin{equation}
	-w^{*}\frac{\pd}{\pd w^{*}}\left(\xp{}(w^*)^{i\nu-\frac{n}{2}}\right)=(w^*)^{i\nu-\frac{n}{2}}
	\label{eqn:wdws}
\end{equation}
which suggests that the insertion of $\xp{}$
corresponds to appending a letter $(-y_0)=(-x_0)$ to the right.  In fact, for the
integrals considered in this paper we can explicitly show that this is correct
on the level of the integral, i.e. we find
\begin{align}
	-w\partial_w\CI\left[\xp{n_+}\xm{n_-+1}\left(\prod_i D_\nu^{a_i}\Ep\right)\Ep^k\right]&=\CI\left[\xp{n_+}\xm{n_-}\left(\prod_i D_\nu^{a_i}\Ep\right)\Ep^k\right],\label{eq:wdw_int}\\
	-w^*\partial_{w^*}\CI\left[\xp{n_++1}\xm{n_-}\left(\prod_i D_\nu^{a_i}\Ep\right)\Ep^k\right]&=\CI\left[\xp{n_+}\xm{n_-}\left(\prod_i D_\nu^{a_i}\Ep\right)\Ep^k\right].
	\label{eq:wdw_ints}
\end{align}
Since the proof of these formulas is not very insightful, the calculation can
be found in \appref{sec:deq_wdw}.  Again comparing with
\eqn{eqn:svhpldiff} this shows that the result of inserting $\xm{}$ into
an integral of type (\ref{eqn:integraltype}) is given by attaching $(-x_0)$
to the left of the result of the integral without insertions, up to possible
constants (which have to be $\zm$-values to preserve uniform
transcendentality).  Similarly, inserting $\xp{}$ in an integrand is given by
attaching $(-x_0)$ to the right of the result of the integral without insertion.
However, note from \eqn{eqn:svhpldiff} that the natural alphabet acting on the
right-hand side of a SVHPL consists of the letters $y_0$ and $y_1$.  Because of
this -- on top of the simple prescription of appending $(-x_0)$ -- we find
additional terms containing $\zm$-values when inserting a factor $\xp{}$ in
the integrand.  These additional terms are exactly those needed to preserve
\eqn{eq:wdw_ints}.  We will comment on these terms in more detail in
\secref{sec:gen_int}.  It should be noted, however, that the complications
beyond simply attaching factors of $(-x_0)$ for insertions of $\chi_\pm$ always
come with $\zm$-values, which in turn means that the $\zm$-free parts are
fully understood from \eqns{eq:wdw_int}{eq:wdw_ints}.

Before closing this section, let us write down specific examples of the
formul\ae{} (\ref{eq:wdw_int}, \ref{eq:wdw_ints}) in the simplest
case
\begin{align}
  \CI[\xp{}]=-\CI[1]x_0\quad\text{and}\quad \CI[\xm{}]=-x_0\CI[1].
  \label{eqn:basisrules1}
\end{align}
The notation in the equation above means that letters $x_0$ are
concatenated to each of the words in $\CI[1]$, using the identification \eqn{eq:rep_xl}. For the first and second part of
\eqn{eqn:basisrules1} we need to append $(-x_0)$ to the right and to the left,
respectively, in order to find from \eqn{eq:res_I1} 
\begin{equation}
  \CI[\xp{}]=-\CL_{00}(-w)-2\CL_{10}(-w)\quad\text{and}\quad \CI[\xm{}]=-\CL_{00}(-w)-2\CL_{01}(-w)\,.
\end{equation}
\Eqn{eqn:basisrules1} can be easily generalized to 
\begin{equation}
 \CI[\xp{n_+}\xm{n_-}]=(-1)^{n_++n_-}x_0^{n_-}\CI[1]\,x_0^{n_+}\,.
 \label{eqn:xpxmsol}
\end{equation}
One might wonder, why for higher orders of the insertion $\xp{}$ there is no
appearance of any $\z$'s which can be attributed to the alphabet $y_0,y_1$.
The na\"ive reason is that the initial integral $\CI[1]$ contains exactly
one letter $x_1$. As powers of $\xp{}$ and $\xm{}$ lead to appending letters
$(-x_0)$ only, there will never be a polylogarithm corresponding to a word with
more than one letter $x_1$.  Comparing with (also higher orders of)
\eqn{eqn:fixpointsol}, one recognizes that any $\zm$-contribution comes with
words containing at least two letters $x_1$. As those can never be created by
appending $x_0$ to $\CI[1]$, there will be no $\zm$-corrections to
\eqn{eqn:xpxmsol}.

\subsection{Basis integrals}
\label{ssec:basisintegrals}

Before proceeding to rules for general integrals of the form
\eqn{eqn:integraltype}, let us consider what we call \textit{basis integrals},
that is those without any insertions of $\xp{}$ and $\xm{}$: 
\begin{equation}
  \CI\left[\big(\prod_i D_\nu^{a_i}\Ep\big)\Ep^k\right]\,.
  \label{eqn:basisintegraltype}
\end{equation}
The residue of those integrals at $i\nu=n/2$ is most efficiently calculated
employing Euler \mbox{$Z$-sums}, a technology which has already been employed
in the context of the Wilson loop OPE in refs.~\cite{Papathanasiou:2014yva, Drummond:2015jea}.
The mechanism works by recognizing that neither $\Ep$ nor
its derivatives have any poles at $i\nu=n/2$.  Therefore any pole at $i\nu=n/2$
with $n>0$ in the basis integrals will be of first order (cf. \eqn{eqn:defI})
and the integral can thus be easily calculated. Evaluating derivatives on $\Ep$
one finds:
\begin{align}
  D_\nu^k\Ep&=
  \psi^{(k)}(1+i\nu+n/2)+(-1)^k\psi^{(k)}(1-i\nu+n/2)\nnl
  D_\nu^k\Ep\Big|_{i\nu=n/2}&=\psi^{(k)}(1+n)+(-1)^k\psi^{(k)}(1)\,.
  \label{eq:DerEPsi}
\end{align}
Employing 
\begin{align}
  \psi^{(m)}(n)=(-1)^{m+1}m!\left(\z(m+1)-Z_{m+1}(n-1)\right),\nnl
  \label{eqn:psiZ}
\end{align}
where by definition
\begin{equation}
	Z_m(n):=\sum\limits_{\ell=1}^n\frac{1}{\ell^m},\mathrm{\quad and \quad}Z_{m_1,\dots,m_k}(n):=\sum\limits_{\ell=1}^n\frac{1}{\ell^{m_1}}Z_{m_2,\dots,m_k}(\ell-1),
	\label{eq:defZsum}
\end{equation}
and generalizing to several powers of the derivative $D_\nu$, yields
\begin{align}
  \Ib[(D_\nu^k\Ep)^q]=\sum_{n=1}^\infty(-1)^kk!\frac{2(-w)^n}{n}\left(Z_{k+1}(n)-(1+(-1)^{k})\zm_{k+1}\right)^q\,.
\end{align}
For several powers of the function $\Ep$ the residue reads
\begin{align}
  \Ib[\Ep^q]=\sum_{n=1}^\infty\frac{2(-w)^n}{n}Z_1(n)^q\,.
  \label{eq:e0_powers}
\end{align}
The resulting $Z$-sums can now be converted into polylogarithms: this is done
with ease employing the identity (for this and the following identities see
\cite{Moch:2001zr}) \footnote{Note that \eqn{eqn:ZtoHPL} only holds if the
index does not contain trailing $0$'s in which case the harmonic polylogarithm
contains logarithmic terms. However, in the integrals we consider this is never
the case.}
\begin{align}
  H_{as}(w)=\sum_{k=1}^{\infty}\frac{w^k}{k^a}Z_s(\ell-1)\,
  \label{eqn:ZtoHPL}
\end{align}
and -- in order to shift the argument of the $Z$-sum to have the appropriate form -- 
\begin{align}
  Z_{m_1,\ldots,m_k}(n+c-1)=Z_{m_1,\ldots,m_k}(n-1)+\sum_{j=0}^{c-1}\frac{1}{(n+j)^{m_1}}Z_{m_2,\ldots,m_k}(n+j-1)\,.
\label{eq:Z_shift}
\end{align}
The identity \eqn{eqn:ZtoHPL} above, however, can be applied to single $Z$-sums
exclusively. Therefore products of $Z$-sums of the same argument have to be
converted into single $Z$-sums, which can be done by recursively using the
stuffle relation 
\begin{align}
  &Z_{m_1,\ldots,m_p}(n)Z_{m'_1,\ldots,m'_q}(n)\nnl
  &\qquad=\sum_{i_1=1}^{n}\frac{1}{i_1^{m_1}}Z_{m_2,\ldots,m_p}(i_1-1)Z_{m'_1,\ldots,m'_q}(i_1-1)\nnl
	&\quad\qquad+\sum_{i_2=1}^{n}\frac{1}{i_2^{m'_1}}Z_{m_1,\ldots,m_p}(i_2-1)Z_{m'_2,\ldots,m'_q}(i_2-1)\nnl
	&\quad\qquad+\sum_{i=1}^{n}\frac{1}{i^{m_1+m'_1}}Z_{m_2,\ldots,m_p}(i-1)Z_{m'_2,\ldots,m'_q}(i-1)\,.
\end{align}

In order to evaluate the full integral we also need to consider the residues at
the double pole $\nu=n=0$. The function $\Ep$ vanishes at
$\nu=n=0$, which lets the residue vanish as soon as any power greater than $1$
of $\Ep$ is present in any basis integral.  As soon as there are, however,
derivatives of $\Ep$ only, there will be a contribution.  The general case can
be calculated by simply expanding the relation
\begin{equation}
	\Iz\left[\big(\prod_i D_\nu^{a_i}\Ep\big)\Ep^k\right]=\left.D_\nu\left(w^{i\nu}\big(\prod_i D_\nu^{a_i}\Ep\big)\Ep^k\right)\right|_{\nu=0},
\end{equation}
from which we find the special cases
\begin{align}
	\Iz[\Ep^q]&\sim\Ep^{q-1}\big|_{\nu=0}=0,\quad\text{for}\,q>1\nnl
  \Iz[D^k_\nu\Ep]&=
  \begin{cases}
    -2(k+1)!\zm_{k+2}& k \quad\text{odd}\\
    -2k!\,H_0(w)\zm_{k+1}&k \quad\text{even}\,
  \end{cases}
  \label{eq:spec_case_res0}
\end{align}
for $k>0$.
This completes our discussion of the basis integrals and we now move on to
explain the calculation of a general integral.

\subsection{General integral}
\label{sec:gen_int}

According to \eqn{eqn:wdw}, adding one power of $\xm{}$ to an integrand
corresponds to prepending a letter $(-x_0)$ to the left for any type of
integrand. Unfortunately, the rules for attaching $x_0$'s to the right 
are not as simple: as already discussed in \subsecref{ssec:Ixpxm} the natural
alphabet to append to the right-hand side of an SVHPL is $y_0$, $y_1$.
Let us show in an explicit example how this can be taken care of:  Assume
we have an integral which evaluates to $\CL_{101}$.  If we insert a factor
$\xp{}$ in the integrand we na\"ively expect to find the result $-\CL_{1010}$.
However, this result violates the relation (\ref{eq:wdw_ints}) since
\begin{equation}
  -w^*\partial_{w^*}(-\CL_{1010})=\CL_{101}+4\zm_3\frac{w^*}{1+w^*},
\end{equation}
as can be easily
checked from \eqn{eq:ex_zeta_svhpl}.  To preserve relation
(\ref{eq:wdw_ints}) the true integral after the insertion of $\xp{}$ will
therefore read $-\CL_{1010}-4\zm_3\CL_{1}$.  To reiterate, when inserting
factors of $\xp{}$ in an integrand, on top of simply appending factors of
$(-x_0)$ to the right, we will find additional contributions containing
$\zm$-values which arise exactly to preserve the differential equations
(\ref{eq:wdw_ints}).
Below we are going to show how to get these seemingly peculiar contributions
under control. We are going to relate residues of different integrals, hereby
tracing back every single integral to one of the basis integrals described in
\subsecref{ssec:basisintegrals}. 

The analysis will be split into a part in which we describe the recursive structure
of the residues at $i\nu=n/2$, followed by a second part, in which we carry out the analysis
of the residue at $\nu=n=0$.

\paragraph{Residue at $i\nu=n/2$.} From the above analysis it became clear that
adjoining any power of $\xm{}$ to an integrand $\CF$ will result in prepending the
appropriate number of letters $x_0$ to the result $\CI[\CF]$.
The only subtlety is the potential appearance of $\zm$-values as integration constants, which, however, only come from the residue at $\nu=n=0$ and will be discussed later.
We can therefore exclude powers of $\xm{}$ in the analysis below and focus on insertions of $\xp{}$. 

In order to derive the relation between different residues at $i\nu=n/2$, let
us get started with the most general integrand not containing $\xm{}$ and
calculate its residue:
\begin{align}
  \Ib&\left[\xp{k+1}(\prod_i D^{a_i}\Ep)\Ep^q\right]=(-2i)\sum\limits_{n=1}^\infty(-1)^n\frac{(-1)^{k+1}}{(k+1)!}D_\nu^{k+1}\left(\frac{w^{i\nu+n/2}}{\nu-\frac{in}{2}}(\prod_i D^{a_i}\Ep)\Ep^q\right)\Bigg|_{\nu=-\frac{i n}{2}}
  \label{eqn:Ibstart}
\end{align}
Aiming for a recursion, we will perform just one of the derivatives $D_\nu$,
yielding
\begin{align}
	&=(-2i)\sum_n(-1)^n\frac{(-1)^{k+1}}{(k+1)k!}D_\nu^k\Bigg(\log w\frac{w^{i\nu+n/2}}{\nu-\frac{in}{2}}\big(\prod_i D_\nu^{a_i}\Ep\big)\Ep^q+i\frac{w^{i\nu+n/2}}{\left(\nu-\frac{in}{2}\right)^2}\big(\prod_i D_\nu^{a_i}\Ep\big)\Ep^q\nnl
       &\hspace{8cm}+\frac{w^{i\nu+n/2}}{\nu-\frac{in}{2}}D_\nu\Big(\big(\prod_i D_\nu^{a_i}\Ep\big)\Ep^q\Big)\Bigg)\Bigg|_{\nu=-\frac{i n}{2}}\,.
\end{align}
The above expression can be rewritten as a sum of integrals evaluated at
$i\nu=n/2$
\begin{align}
       &=\frac{1}{(k+1)}\Bigg(-H_0(w)\Ib\left[\xp{k}(\prod_i D_\nu^{a_i}\Ep)\Ep^q\right]-\Ib\left[\xp{k}\xm{}(\prod_i D_\nu^{a_i}\Ep)\Ep^q\right]\nnl
       &\hspace{8cm}-\Ib\left[\xp{k}D_\nu\Big((\prod_iD_\nu^{a_i}\Ep)\Ep^q\Big)\right]\Bigg)\,.
  \label{eqn:resb}
\end{align}
The integrands on the right-hand side of the above equation have less powers of
$\xp{}$ compared to the integrand we started with in \eqn{eqn:Ibstart}.
Therefore we can successively get rid of \emph{all} powers of $\xp{}$ by
trading them for easier integrals. At the end of the recursion we are left with
just the basis integrals discussed \subsecref{ssec:basisintegrals} above and
insertions of $\xm{}$, which we understand.

\paragraph{Residue at $\nu=n=0$.} For the second residue contributing to the
calculation of the remainder function, the analysis is slightly simpler, but
follows a similar pattern. Again, we start from the most general integral,
however, this time we have to consider powers of $\xm{}$ as well, as
those contribute in the same way as $\xp{}$: for $n=0$ both of them reduce to
$i/\nu$ (cf. \eqns{eqn:xpxm}{eqn:definitions}). Starting from an
integral with $k$ powers of either $\xp{}$ or $\xm{}$, we find
\begin{align}
  \Iz\left[\left(\frac{i}{\nu}\right)^k(\prod_i D^{a_i}\Ep)\Ep^q\right]
  =(-1)^k\frac{D_\nu^{k+1}}{(k+1)!}\left(w^{i\nu}(\prod_i D^{a_i}\Ep)\Ep^q\right)\Bigg|_{\nu=0}\,.
\end{align}
Performing again just one of the derivatives $D_\nu$, yields
\begin{align}
		     &=(-1)^k\frac{D_\nu^{k}}{(k+1)!}\Bigg(w^{i\nu}\log w(\prod_i D^{a_i}\Ep)\Ep^q
  +w^{i\nu}D_\nu\Big(\big(\prod_i D_\nu^{a_i}\Ep\big)\Ep^q\Big)\Bigg)\Bigg|_{\nu=0}\,,
\end{align}
which is again the sum of several different residues:
\begin{align}
  =\frac{1}{(k+1)}\Bigg(-H_0(w)\Iz\left[\left(\frac{i}{\nu}\right)^{k-1}(\prod_i D_\nu^{a_i}\Ep)\Ep^q\right]-
    \Iz\left[\left(\frac{i}{\nu}\right)^{k-1}D_\nu\Big(\big(\prod_i D_\nu^{a_i}\Ep\big)\Ep^q\Big)\right]\Bigg)\,.
  \label{eqn:resz}
\end{align}
Similar to the recursive formalism used above for the contribution from
$i\nu=n/2$, one can thus recursively trace back any situation to the residues
at $\nu=n=0$ of the basis integrals. 

\subsection{Example: $\CI[(D_\nu^2 E)E]$}

In order to illustrate the above method, let us consider a simple but
nontrivial example. The integral $\CI[(D_\nu^2 E)E]$, which appears at four
loops in NLLA, can be expanded as  
\begin{align}
  \CI[(D_\nu^2 E)E]&=
  \CI\Big[\frac{\xm{4}}{2} - \frac{\xm{3} \xp{}}{2} - \frac{\xm{} \xp{3}}{2}+ \frac{\xp{4}}{2} + \xm{3} \Ep - \xp{3} \Ep \nnl
 &\qquad\qquad\quad + \frac{\xm{} D_\nu^2\Ep}{2} - \frac{\xp{} D_\nu^2\Ep}{2} + \Ep D_\nu^2\Ep\Big]\,.
\end{align}
The integrals, which contain powers of $\xp{}$ and $\xm{}$ only, can be
trivially solved using \eqn{eqn:xpxmsol},
\begin{align}
  \CI\Big[\frac{\xm{4}}{2} - \frac{\xm{3} \xp{}}{2} - \frac{\xm{} \xp{3}}{2}+ \frac{\xp{4}}{2}\Big]=\CL_{00001}-\CL_{00010}-\CL_{01000}+\CL_{00001}\,,
\end{align}
where we suppressed the arguments $w$ of the SVHPLs. Using the direct
methods described in \subsecref{ssec:basisintegrals} in order to derive the basis integrals
\begin{align}
	\CI[\Ep]&=\left(\Ib[\Ep]+\Iz[\Ep]\right)\Big|_{H\rightarrow\CL}=2(\CL_{01}+\CL_{11})\,,\nnl
	\CI[D_\nu^2\Ep]&=\left(\Ib[D_\nu^2\Ep]+\Iz[D_\nu^2\Ep]\right)\Big|_{H\rightarrow\CL}=4\big(\CL_{0001}+\CL_{1001}-2\zm_3\CL_1\big)-4\zm_3\CL_0\,,\nnl
  \CI[\Ep D_\nu^2\Ep]&=4\big(\CL_{00001}+\CL_{00011}+\CL_{01001}+\CL_{10001}+\CL_{10011}+\CL_{11001}-2\zm_3(\CL_{01}+\CL_{11})\big)\,.
\end{align}
one can now tackle the other integrals. By just attaching the appropriate
number of letters $(-x_0)$ from the left matching the power of $\xm{}$, one finds
immediately
\begin{align}
	\CI[\xm{3}\Ep]&=(-x_0)^3\left(\Ib[\Ep]\Big|_{H\rightarrow\CL}\right)+\Iz[\xm{3}\Ep]\Big|_{H\rightarrow \CL}\nnl
		&=-2(\CL_{00001}+\CL_{00011})+2\CL_{00}\zm_3+2\zm_5\,,\nnl
		\CI[\xm{}D^2_\nu\Ep]&=-x_0\left(\Ib[D_\nu^2\Ep]\Big|_{H\rightarrow\CL}\right)+\Iz[\xm{}D^2_\nu\Ep]\Big|_{H\rightarrow\CL}\nnl
	 &=-4\Big(\CL_{00001}+\CL_{01001}-2\zm_3\CL_{01}\Big)+4\Big(\zm_3\CL_{00}+6\zm_5\Big)\nnl
	 &=-4(\CL_{00001}+\CL_{01001}-\zm_3(\CL_{00}+2\CL_{01})-6\zm_5)\,.
\end{align}
Let us now evaluate the integral $\CI[\xp{3}\Ep]$. In order to get started, let
us consider the residue at $i\nu=n/2$, which we will rewrite using
\eqn{eqn:resb}:
\begin{align}
	\Ib[\xp{3}\Ep]&=\frac{1}{3}\left(-H_0(w)\Ib[\xp{2}\Ep]-\Ib[\xp{2}\xm{}\Ep]-\Ib[\xp{2}D_\nu\Ep] \right)\,.
\end{align}
Naturally, one would now reduce $\Ib[\xp{2}\Ep]$ in the same manner and finally
continue with the integral $\Ib[\xp{}\Ep]$. After evaluating everything, one
finds
\begin{align}
  \Ib[\xp{3}\Ep]&=-2(H_{01000}+H_{11000})+4\zm_3 H_{10}\,,\nnl
  \Ib[\xp{2}\Ep]&=2(H_{0100}+H_{1100})-4\zm_3 H_1\,,\nnl
  \Ib[\xp{}\Ep]&=-2(H_{010}+H_{110})\,.
\end{align}
What remains to be done is to calculate the contributions from the residue at
$\nu=n=0$ for the above integrals. This can be done in (almost) complete
analogy to the considerations above by using \eqn{eqn:resz}. One finds:
\begin{align}
  \Iz[\xp{3}\Ep]&=2\zm_3 H_{00}+2\zm_5\,,\nnl
  \Iz[\xp{2}\Ep]&=-2\zm_3 H_{0}\,,\nnl
  \Iz[\xp{}\Ep]&=2\zm_3\,.
\end{align}
Adding up all contributions and using the replacement rule (\ref{eq:res_presc}), one finally finds:
\begin{align}
  \CI[(D_\nu^2 E)E]&=
\CL_{0 0 0 0 1} + \CL_{0 0 0 1 0} + 2 \CL_{0 0 0 1 1} + 
 \CL_{0 1 0 0 0} + 2 \CL_{0 1 0 0 1} \nnl
 &\quad+ \CL_{1 0 0 0 0} + 
 4 \CL_{1 0 0 0 1} + 2 \CL_{1 0 0 1 0} + 
 4 \CL_{1 0 0 1 1} + 2 \CL_{1 1 0 0 0} \nnl
 &\quad+ 4 \CL_{1 1 0 0 1} - 4 \zm_3\big(\CL_{0 1} + 2 \CL_{1 0}+2 \CL_{1 1}\big)\,.
\end{align}
As should be clear from our example, these calculations will become tedious for
high weights if carried out by hand, but are very simple to implement on a
computer. In particular, given the recursive structure there are very
few integrals to be determined at all. 


\section{NMHV}
\label{sec:nmhv}

While the discussion in the previous sections was tailored to obtaining the
six-point remainder function $R_6^\MHV$, we would like to apply our method to
the NMHV sector. Following the setup described in \rcite{Dixon:2014iba}, we
flip the helicity of the external gluon 4 compared to the MHV-situation in
order to obtain the helicity configuration $(++--+-)$.\footnote{Note that in
the high-energy limit, only the helicities of the external gluons $4$ and $5$
can be flipped, since helicity is conserved along the high-energy lines.}
Naturally, the kinematical setup remains the same as described in
\eqns{eq:mrl_crs}{eq:mrl_red_crs}. Below we are going to relate the integrals
appearing in the NMHV sector to those which have already been solved for the
MHV sector above. This idea has already been proven useful in many situations
\cite{Dixon:2012yy,Lipatov:2012gk,Dixon:2014iba} and has been discussed therein
thoroughly. Therefore, we sketch the formalism only briefly and point out where
particular adjustments to our setup are necessary.  The remainder function
$R_6^\NMHV$ is defined via
\begin{equation}
  R_6^\NMHV=\frac{A_\NMHV}{A_\text{BDS}}=\CP_\NMHV\times e^{R_6^\MHV} 
\end{equation}
and can be determined by evaluating the following integral
\cite{Dixon:2014iba}:
\begin{align}
  \CP_\NMHV\times e^{R+i\pi\delta}|_\MRK=&\cos \pi\omm_{ab}-i\frac{a}{2}\sum_{n=-\infty}^{\infty}(-1)^n
  \left(\frac{w}{w^*}\right)^{\frac{n}{2}}\int_{-\infty}^\infty\frac{\dd\nu}{\left(i\nu+\frac{n}{2}\right)^2}\,|w|^{2i\nu}\Phi_\reg^\NMHV(\nu,n)\nnl
  &\qquad\qquad\times\,\exp\left[-\omm(\nu,n)\left(\log(1-u_1)+i\pi+\frac{1}{2}\log\frac{|w|^2}{|1+w|^4}\right)\right]\,.
  \label{eqn:masternmhv}
\end{align}
The above formula differs from \eqn{eqn:master} in two respects: the factor
determining the residue in each integral is altered via
\begin{align}
  \frac{1}{\nu^2+\frac{n^2}{4}}=-\xp{}\xm{}\to-\frac{1}{\left(i\nu+\frac{n}{2}\right)^2}=-\xm{2}
\end{align}
and the MHV impact factor is replaced by $\Phi_\reg^\NMHV$, the impact factor for NMHV.
It can be can be obtained at any loop order by supplementing
$\Phi^\MHV_\reg$ with a suitable correction factor, which, in turn, can be
inferred and derived from integrability \cite{Basso:2014pla, Dixon:2014iba}.

Having all the ingredients for the integrand available, we are left with
evaluating the integral. Fortunately one does not explicitly need to solve this
particular integral: defining
\begin{equation}
  \CI^\NMHV[\CF(\nu,n)]=\frac{1}{\pi}\sum\limits_{n=-\infty}^{\infty}(-1)^n\left(\frac{w}{w^*}\right)^{\frac{n}{2}}\int_{-\infty}^\infty\frac{\dd\nu}{(i \nu+\frac{n}{2})^2}\,|w|^{2i\nu}\,\CF(\nu,n) \,.
  \label{eqn:defINMHV}
\end{equation}
one easily finds from \eqns{eqn:wdw}{eqn:wdws} that
\begin{equation}
  w\frac{\pd}{\pd w}\CI^\NMHV[\CF(\nu,n)]=-w^*\frac{\pd}{\pd w^*}\CI^\MHV[\CF(\nu,n)]\,,
\end{equation}
where $\CI^\MHV$ refers to the integral defined in \eqn{eqn:defI}. Derivatives
with respect to the variable $w^*$, however, can be easily performed using
\eqn{eqn:svhpldiff} after taking the identification $(z,z^*)=(-w,-w^*)$
into account.

The ratio function $\CP_\NMHV$ can be expanded in the loop order as 
\begin{align}
  \CP_\NMHV|_\MRK=1+\sum_{\ell=2}^{\infty}\sum_{n=0}^{\ell-1}a^\ell\log^n(1-u_1)
  &\Bigg\lbrace\frac{1}{1+w^*}\big[p_n^{(\ell)}(w,w^*)+2\pi i q_n^{(\ell)}(w,w^*)\big]\nnl
  +\frac{w^*}{1+w^*}\big[p_n^{(\ell)}(w,w^*)&+2\pi i q_n^{(\ell)}(w,w^*)\big]\Bigg|_{(w,w^*)\to(1/w,1/{w^*})}\Bigg\rbrace\,.
  \label{eqn:PNMHVexpansion}
\end{align}
Starting from \eqn{eqn:masternmhv}, one will have to solve the integral, plug
in the appropriate functions $g^{(\ell)}_n$ and $h^{(\ell)}_n$ for $R_6^\MHV$
and bring the result into the form (\ref{eqn:PNMHVexpansion}) to determine the
functions $p^{(\ell)}_n$ and $q^{(\ell)}_n$. Following this method, we can confirm all previous
results in refs.~\cite{Lipatov:2012gk,Dixon:2014iba}. Higher orders can be
obtained with the same ease and comparable computational effort as in the MHV
case.


\section{Applications}
\label{sec:applications}
Using our algorithm as described above, we can straightforwardly produce new
data for a high number of loops.  As specific applications of our technique, we
therefore present plots of the imaginary parts $g^{(\ell)}_{\ell-4}$ at
N$^3$LLA up to ten loops and $g^{(\ell)}_{\ell-5}$ up to nine loops, as well as
new expressions for the remainder function in the collinear-Regge limit.

\subsection{N$^3$LLA/N$^4$LLA}
\label{sec:app_hl}
\begin{figure}[t]
	\centering
	\includegraphics[scale=1.3]{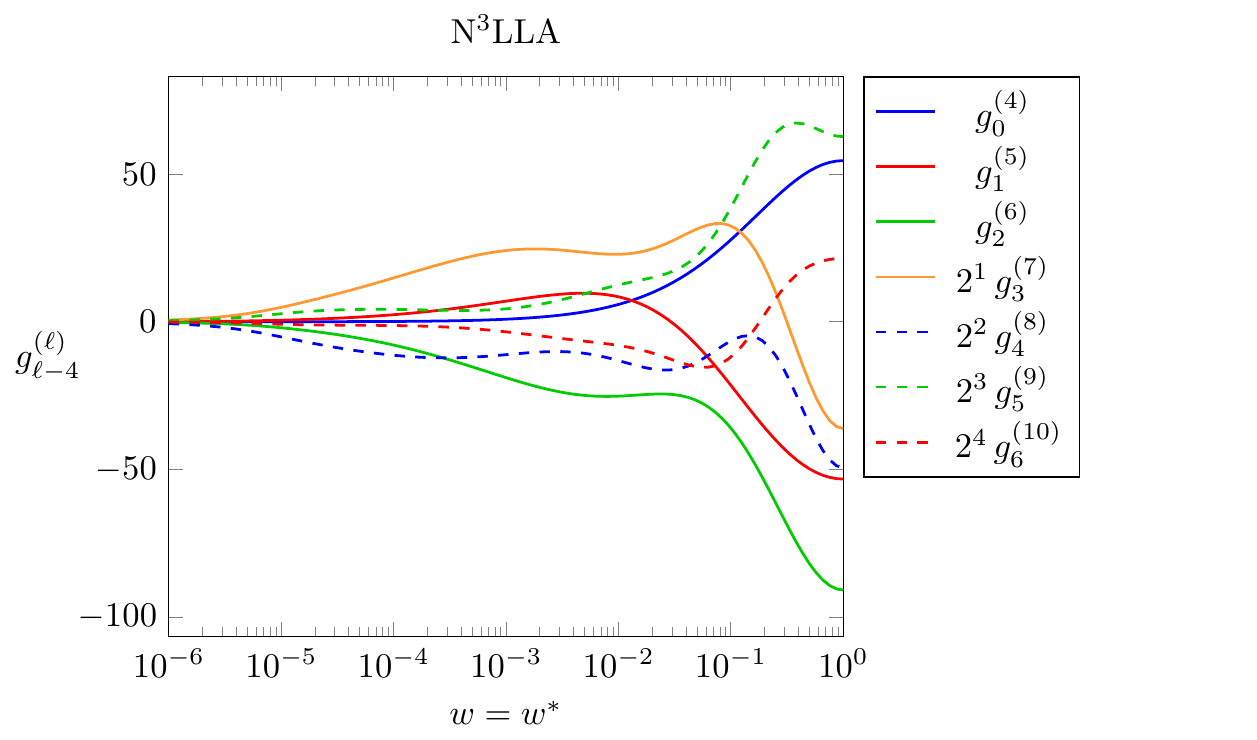}
	\caption{Behavior of the N$^3$LLA-functions $g^{(\ell)}_{\ell-4}$ along
	the line $w=w^*$. Note that some of the functions have been rescaled
      for convenience.}
	\label{fig:n3lla}
\end{figure}

\begin{figure}[t]
	\centering
	\includegraphics[scale=1.3]{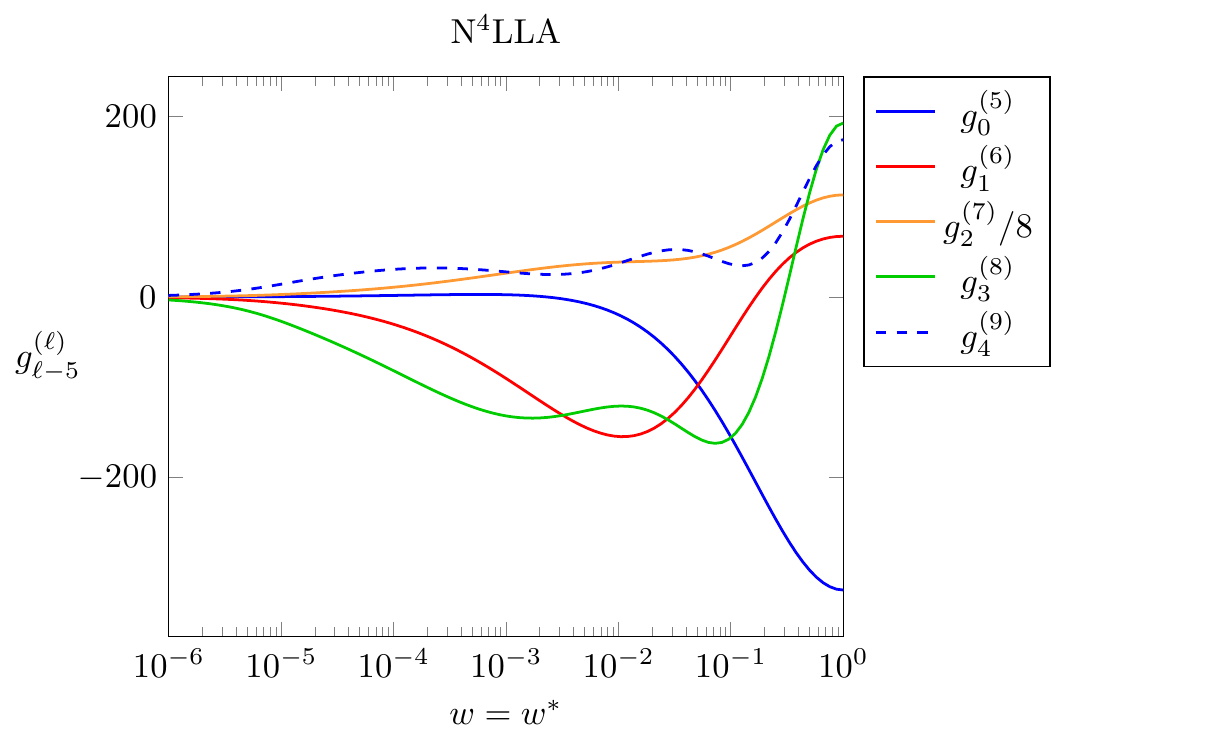}
	\caption{Behavior of the N$^4$LLA-functions $g^{(\ell)}_{\ell-5}$ along
	the line $w=w^*$. Note that some of the functions have been rescaled
      for convenience.}
	\label{fig:n4lla}
\end{figure}

As the simplest application of our algorithm, we investigate the behavior of
the previously-unknown functions $g^{(\ell)}_{\ell-4}$, $g^{(\ell)}_{\ell-5}$
beyond five loops.  In figures \ref{fig:n3lla} and \ref{fig:n4lla} we plot
these functions along the line $w=w^*$ up to ten loops for N$^3$LLA and nine
loops for N$^4$LLA.  To produce these plots, we have used GiNaC
\cite{Bauer:2000cp} for the numerical evaluation of the HPLs.

Let us briefly comment on the performance of our algorithm.  Using the most
straightforward implementation of our algorithm in \textrm{Mathematica} on a
standard laptop, we obtain the expressions up to ten loops for LLA and NLLA in
about one minute each and for N$^3$LLA in about ten minutes.  We furthermore
determined the full remainder function in the high-energy limit up to seven loops
by calculating $g^{(7)}_k$ for $k=0,\dots,6$, which took of the order of $100$
minutes, due to the high logarithmic degree for $k=0$.  It should be possible
to reduce these numbers a lot by a more sophisticated implementation.  However,
usually we do not stop at the expression in terms of SVHPLs, but rather want to
compute series expansions or generate data points such as those shown in
figures \ref{fig:n3lla}, \ref{fig:n4lla}.  To do so, we need to know the
expressions of the SVHPLs in terms of HPLs up to high weights, which, in turn,
requires the solution of the single-valuedness condition in \rcite{Brown2004527} for
high weights.  Both solving the single-valuedness condition as well as series expansions
or numerical evaluations of the HPLs become time-consuming for high weights.
Also taking into account that the number of terms appearing in the
$g^{(\ell)}_n$ grows with the logarithmic degree, we think that it should be
possible to obtain results a couple of orders beyond the results presented
here, but not much further.

\subsection{Collinear-Regge limit}
\label{sec:app_crl}
As described in \eqn{eq:mrl_red_crs} the remaining kinematical freedom in the
multi-Regge limit is described by the complex variable $w$.  We therefore have
the freedom of taking another limit on top of the multi-Regge limit,
\begin{equation}
	|w|\rightarrow 0\,,
	\label{eq:def_col_regge}
\end{equation}
which describes the collinear-Regge regime.  In this limit, we in principle
have two different kinds of large logarithms, namely $\log(1-u_1)$ and
$\log|w|$. However, the remainder function organizes itself effectively as a
function of a single variable,
\begin{equation}
	x:=a\log(1-u_1)\log|w|\,.
	\label{eq:def_colregx}
\end{equation}
Indeed, the leading contributions, i.e.~those with the maximal powers of
$\log|w|$ at a given loop order, in the LLA collinear-Regge remainder function
can be explicitly resummed in a very simple form
\begin{equation}
	\left.R_{6}^{\MHV}\right|_{\mathrm{LLA, coll.}}=i \pi a(w+w^*)\left(1-I_0(2\sqrt{x})\right)\,,
	\label{eq:r6_dlla}
\end{equation}
where $I_0(x)$ is a modified Bessel function, cf.\ \cite{Bartels:2011xy}.
These leading contributions are also referred to as the
double-leading-logarithmic approximation (DLLA), with an obvious generalization
beyond LLA.  Note that the prefactor $(w+w^*)$ in \eqn{eq:r6_dlla} is
necessary to ensure that the remainder function vanishes in the collinear
limit.  A closed formula for all subleading contributions was presented in
\rcite{Pennington:2012zj}.\par

Using our algorithm, we can generate data up to a high number of loops.
Finding the leading contributions in the collinear-Regge limit is then simply a
matter of expanding the appearing SVHPLs in the limit $|w|\rightarrow 0$.  This
allows us to write down conjectures for the leading contributions to the
imaginary part of the remainder function in the collinear-Regge limit for
N$^{(1,2,3)}$LLA:
\begin{alignat}{3}
	& \left.\Im R^\MHV_6\right|_{\mathrm{DNLLA}} && =2\pi a^2\log|w|(w+w^*) && \left(-\frac{1}{2}I_0(2\sqrt{x})-\frac{1}{2}\pd_xI_0(2\sqrt{x})\right)\,,\nnl
	& \left.\Im R^\MHV_6\right|_{\mathrm{DN^2LLA}}&&=2\pi a^3\log^2|w|(w+w^*) && \left[\left(-\frac{1}{4}I_0(2\sqrt{x})-\pd_xI_0(2\sqrt{x})-\frac{1}{2}\pd^2_xI_0(2\sqrt{x})\right)\right.\nnl
	& && &&\left.+\zeta_2\left(\frac{3}{2}I_0(2\sqrt{x})-2\pd_xI_0(2\sqrt{x})+\frac{3}{2}\pd^2_xI_0(2\sqrt{x})\right)\right]\,,\nnl
	& \left.\Im R^\MHV_6\right|_{\mathrm{DN^3LLA}}&&=2\pi a^4\log^3|w|(w+w^*) && \left[\left(-\frac{1}{12}I_0(2\sqrt{x})-\frac{3}{4}\pd_xI_0(2\sqrt{x})-\frac{3}{2}\pd^2_xI_0(2\sqrt{x})\right.\right.\nnl
	& && &&\left.-\frac{1}{2}\pd^3_xI_0(2\sqrt{x})\right)+\zeta_2\left(\frac{3}{2}I_0(2\sqrt{x})-\frac{1}{2}\pd_xI_0(2\sqrt{x})\right.\nnl
	& && &&\left.\left.-\frac{1}{2}\pd^2_xI_0(2\sqrt{x})+\frac{3}{2}\pd^3_xI_0(2\sqrt{x})\right)+\zeta_3\pd^2_xI_0(2\sqrt{x})\right]\,.
	\label{eq:conj_col_regge}
\end{alignat}

Each of these relations was checked up to ten loops.  Note that we refrain from
spelling out the real parts which can be obtained from the imaginary parts, as
explained before.  In \eqn{eq:conj_col_regge} we have chosen to write all
relations in terms of $I_0(2\sqrt{x})$ and derivatives thereof.  Whether a
simpler form can be found by using the numerous identities among the functions
$I_n(2\sqrt{x})$ is left as an open question for future research.

\section{Proof of Pennington's formula}
\label{sec:proof}

In \cite{Pennington:2012zj} Pennington put forward a simple formula for the
remainder function in LLA in momentum space.  This formula was stated as a
conjecture, based on the available data.  In this section we proof Pennington's
formula using the techniques explained in the earlier sections.

\subsection{Pennington's formula}

We begin by spelling out the formula for the six-gluon MHV remainder function
in LLA as described in \rcite{Pennington:2012zj},
\begin{equation}
	\left.R_6^\MHV\right|_{\mathrm{LLA}}=\frac{2\pi i}{\log(1-u_1)}\left(\mathcal{X} \mathcal{Z}^{\mathrm{MHV}}-\frac{1}{2}x_1\eta\right),
	\label{eq:rlla}
\end{equation}
where $\eta=a\log(1-u_1)$ and
\begin{align}
	\mathcal{X}&=e^{\frac{1}{2}x_0\eta}\left[1-x_1\left(\frac{e^{x_0\eta}-1}{x_0}\right)\right]^{-1},\\
	\mathcal{Z}^{\mathrm{MHV}}&=\frac{1}{2}\sum\limits_{k=1}^{\infty}\left(x_1\sum\limits_{n=0}^{k-1}(-1)^nx_0^{k-n-1}\sum\limits_{m=0}^n\frac{2^{2m-k+1}}{(k-m-1)!}\mathfrak{Z}(n,m)\right)\eta^k.
\end{align}
The coefficients $\mathfrak{Z}(n,m)$ are generated by the series
\begin{equation}
	\exp\left[y\sum\limits_{k=1}^{\infty}\zm_{2k+1}x^{2k+1}\right]=:\sum\limits_{n=0}^\infty\sum\limits_{m=0}^{\infty}\mathfrak{Z}(n,m)x^ny^m.
	\label{eq:gen_ser_z}
\end{equation}
The above formula can in principle be expanded to any order in the coupling
constant $a$ and was checked to $14$ loops in \rcite{Pennington:2012zj}.

\subsection{Strategy of the proof}

We will prove \eqn{eq:rlla} by direct evaluation of the dispersion relation
\eqn{eqn:master}.  By expanding the dispersion relation \eqn{eqn:master} it
is easy to see that the $\ell$-loop contribution in LLA is given by
\begin{equation}
	g^{(\ell)}_{\ell-1}=\frac{1}{4(\ell-1)!}\CI\left[(E^{(0)})^{\ell-1}\right],
	\label{eq:gl_int}
\end{equation}
while the real part vanishes in LLA, $h^{(\ell)}_{\ell-1}=0$.
Using relation (\ref{eq:e0_split}) the $\ell$-loop contribution to the remainder function can be written as
\begin{equation}
	g^{(\ell)}_{\ell-1}=\frac{1}{4(\ell-1)!}\sum\limits_{k=0}^{\ell-1}\binom{\ell-1}{k}\left(-\frac{1}{2}\right)^k\sum\limits_{m=0}^k(-1)^{k-m}\binom{k}{m}\CI\left[\xp{m}\xm{k-m}\Ep^{\ell-1-k}\right].
	\label{eq:gl_expand}
\end{equation}
As described in \secref{sec:solution} we understand the
insertion of the factors $\chi_\pm$ in the integrand, so we only need to
calculate the basis integrals $\CI\left[\Ep^k\right]$.
These were already considered in \eqn{eq:e0_powers} and we can directly rewrite them in HPLs,
\begin{align}
	\Ib\left[\Ep^k\right]&=(-2i)\sum\limits_{n=1}^\infty\frac{(-w)^n}{-in}Z_1(n)^k=2\sum\limits_{n=1}^\infty\frac{(-w)^n}{n}\sum\limits_{l=1}^k\sum\limits_{\al\in T(l,k)}\frac{k!}{\prod \al_i!}Z_{\al_1,\dots,\al_l}(n)\label{eq:LLA_basis_fl}\\
	&=2\sum\limits_{l=1}^k\sum\limits_{\al\in T(l,k)}\frac{k!}{\prod \al_i!}\left(H_{1,\al_1,\dots,\al_l}(-w)+H_{\al_1+1,\dots,\al_l}(-w)\right),\label{eq:LLA_basis_hpl}
\end{align}
where the first step follows from a simple inductive proof and the second step
uses the definition of the harmonic polylogarithms in terms of $Z$-sums
(\ref{eqn:ZtoHPL}), as well as relation (\ref{eq:Z_shift}).  Furthermore,
$T(l,k)$ is the set of $l$-tuples of non-negative integers such that their sum
is $k$.  Note that the result is written in the collapsed notation for HPLs.
This allows us to immediately write down the full result via the replacement
rule (\ref{eq:res_presc}),
\begin{equation}
	\CI\left[\Ep^k\right]=2\sum\limits_{l=1}^k\sum\limits_{\al\in T(l,k)}\frac{k!}{\prod \al_i!}\left(\mathcal{L}_{1,\al_1,\dots,\al_l}+\mathcal{L}_{\al_1+1,\dots,\al_l}\right),
	\label{eq:basis_int_LLA}
\end{equation}
where we have used that the residues at $\nu=n=0$ vanish for these basis
integrals, see \eqn{eq:spec_case_res0}.  Together with the result for $k=0$,
\eqn{eq:res_I1}, we have determined all necessary integrals.  The proof now
proceeds as follows.  In the next section, we will first prove \eqn{eq:rlla}
for the $\zm$-free part, for which the insertion of $\chi_\pm$ is well
understood.  In \secref{sec:zetapart} we will then examine the appearance
of $\zm$-values and show that for any combination of $\zm$'s the direct
calculation reproduces Pennington's result. 

\subsection[Proof of the $\zm$-free part]{Proof of the \texorpdfstring{$\boldsymbol{\zm}$}{zeta}-free part}
As described in \secref{sec:solution} inserting factors $\chi_\pm$ in the
integrand simply appends or prepends factors of $(-x_0)$, neglecting potential
$\zm$'s for the moment.  We can then immediately write down the full
expression for the remainder function as
\begin{align}
	\frac{1}{2\pi i}\log(1-u_1) & R_{6}^\MHV\Big|_{\zfree}=\sum\limits_{\ell=1}^\infty\eta^\ell g^{(\ell)}_{\ell-1}-\eta g^{(1)}_0\nnl
	&=\sum\limits_{\ell=1}^\infty\sum\limits_{k=0}^{\ell-1}\sum\limits_{m=0}^k\frac{(-1)^{m}}{2^k}\frac{\eta^\ell}{4(\ell-1)!}\binom{\ell-1}{k}\binom{k}{m}\CI\left[\xp{m}\xm{k-m}\Ep^{\ell-1-k}\right]-\frac{\eta}{4}(x_0+2x_1)\nnl
	&=\dots=\sum\limits_{m=0}^\infty\sum\limits_{k=0}^\infty\sum\limits_{\ell=0}^\infty\frac{(-1)^m}{2^{k+m}}\frac{\eta^{\ell+k+m+1}}{4\,\ell!\,m!\,k!}\CI\left[\xp{m}\xm{k}\Ep^\ell\right]-\frac{\eta}{4}(x_0+2x_1)\label{eq:r6_zfree_firstfactor}\\
	&=\frac{\eta}{4}e^{-\frac{\eta}{2}x_0}\left(\sum\limits_{\ell=0}^\infty\frac{\eta^\ell}{\ell!}\CI\left[\Ep^\ell\right]\right)e^{\frac{\eta}{2}x_0}-\frac{\eta}{4}(x_0+2x_1),\label{eq:r6_zfree_factor}
\end{align}
where in the intermediate steps we have only relabeled variables and switched
orders of summation, which is tedious but straightforward.  The last expression
in \eqn{eq:r6_zfree_factor} is already nicely factorized, we only have to
rewrite the factor
\begin{equation}
	\sum\limits_{\ell=0}^\infty\frac{\eta^\ell}{\ell!}\CI\left[\Ep^\ell\right],
\end{equation}
whose summands we have already determined in \eqn{eq:basis_int_LLA}.  To find a
simpler representation let us write down the coefficient of an arbitrary SVHPL
$c_{n_1,\dots,n_m}\mathcal{L}_{n_1,\dots,n_m}$ (in collapsed notation)
appearing in \eqn{eq:basis_int_LLA} which is given by
\begin{equation}
	c_{n_1,\dots,n_m}=\frac{2}{\eta}\frac{\eta^{\sum n_i}}{(n_1-1)!\,n_2!\cdots n_m!},
	\label{eq:coeff_epsi}
\end{equation}
as is easily checked.\footnote{The simplest way to obtain
\eqn{eq:coeff_epsi} is to see that in \eqn{eq:basis_int_LLA} $\ell$
determines the number of $1$s, while $k$ determines the length of the SVHPL.
Writing down the two possible cases for $n_1=1$, $n_1>1$ immediately leads to our
expression.} This, however, is exactly the coefficient one obtains for the
expansion of the compact expression
\begin{equation}
	\sum\limits_{\ell=0}^\infty\frac{\eta^\ell}{\ell!}\CI\left[\Ep^\ell\right]=2 e^{x_0\eta}\left[1-x_1\left(\frac{e^{x_0\eta}-1}{x_0}\right)\right]^{-1}x_1+x_0,
\end{equation}
where the last $x_0$ arises from the $n=0$ residue of $\CI[1]$.
Plugging this result in the remainder function \eqn{eq:r6_zfree_factor} we
obtain the expression written down in \cite{Pennington:2012zj}.

\subsection[Including the $\zm$'s]{Including the \texorpdfstring{$\boldsymbol{\zm}$}{zeta}'s}
\label{sec:zetapart}

Let us now include the $\zm$'s.  To do so, we will determine the resummed
remainder function for an arbitrary combination of $\zm$'s and show the
equivalence with the result of \cite{Pennington:2012zj}.  Since the basis
integrals $\CI\left[\Ep^k\right]$ never contain any $\zm$'s by themselves,
$\zm$-values are only created through insertions of $\xp{}$,
$\xm{}$.\footnote{Here we see another great benefit of working on the level of
  HPLs using the replacement $H\rightarrow\CL$: Any term containing $\zm$'s
will explicitly appear in the final answer, while in the original formulation
$\zm$-terms can potentially be absorbed into the definition of SVHPLs.} We are
therefore led to consider the integrals
\begin{equation}
	\Ib\left[\xp{m}\xm{k}\Ep^\ell\right]+\Iz\left[\xp{m}\xm{k}\Ep^\ell\right],
\end{equation}
which we will abbreviate as $\Ibz\left[\xp{m}\xm{k}\Ep^\ell\right]$.  Let us
first consider the case of a single $\zm$ as a warm-up.  By directly evaluating
the residues we find 
\begin{align}
	\Ibz\left[\xm{k}\xp{m}\Ep^\ell\right]_{\zeta_{a+1}}&=(-2i)i^{m+k}\sum\limits_{n=1}^\infty(-1)^n\frac{i^m}{m!}D_\nu^m\left(\frac{w^{i\nu+\frac{n}{2}}}{\left(\nu-i\frac{n}{2}\right)^{k+1}}\Ep^\ell\right)\Bigg|_{\nu=-i\frac{n}{2}}\nnl
	&\quad\quad\quad+(-i)i^{m+k}\frac{i^{m+k+1}}{(m+k+1)!}D_\nu^{m+k+1}\left(w^{i\nu}\Ep^\ell\right)\Bigg|_{\nu=0}\nnl
	&=i^{a}\ell \frac{(D_\nu^a\Ep)\big|_{\nu=-i\frac{n}{2}}}{a!}\Ib\left[\xm{k}\xp{m-a} \Ep^{\ell-1}\right]_\zfree\nnl
	&\quad\quad\quad+i^{a}\ell \frac{(D_\nu^a\Ep)\big|_{\nu=0}}{a!}\Iz\left[\xm{k}\xp{m-a} \Ep^{\ell-1}\right]_\zfree \nnl
        &=(-2\ell)\zm_{a+1}\Ibz\left[\xm{k}\xp{m-a} \Ep^{\ell-1}\right]_\zfree.
	\label{eq:single_zeta}
\end{align}
In the resulting expression we project the
remaining integral to the $\zm$-free part because we are only interested in the
piece proportional to $\zm_{a+1}$, the neglected terms do, however, contribute
to other $\zm$-structures.
Similarly, in going from the second to third line in \eqn{eq:single_zeta} we have neglected a Euler sum in the derivative $(D_\nu^a \Ep)\big|_{\nu=-i\frac{n}{2}}$, which does not contain any $\zm$'s.
Note that the next-to-last step in \eqn{eq:single_zeta} also shows that only odd $\zm$'s can appear, since even
$\zm$'s never appear in the derivatives of $E_\psi$ (cf.
\eqns{eq:DerEPsi}{eqn:psiZ}).  The step to an arbitrary combination of $\zm$'s
is now straightforward, we only have to be careful about combinatorial factors.
We find
\begin{equation}
	\Ibz\left[\xm{k}\xp{m}\Ep^\ell\right]_{\prod\limits_i\zm_{a_i+1}}
	=(-2)^{n_z}\frac{\ell!}{(\ell-n_z)!}
	  \frac{\prod\limits_i^{n_z}\zm_{a_i+1}}{\prod\limits_{a_i}n_{a_i}!}
	  \Ibz\left[\xm{k}\xp{m-\sum a_i}\Ep^{\ell-n_z}\right]_\zfree,
\label{eq:zeta_rec}
\end{equation}
where $n_z$ is the total number of $\zm$'s and $n_{a_i}$ counts the
multiplicity of $\zm_{a_i+1}$. Using this result, we can immediately calculate the remainder
function proportional to this $\zm$-structure starting from
\eqn{eq:r6_zfree_firstfactor}.  For an arbitrary product of $\zm$'s,
$\prod\limits_i^{n_z}\zm_{a_i+1}$, we need $\ell\geq n_z$, $m\geq\sum\limits_i
a_i$, so we obtain
\begin{align}
	\frac{1}{2\pi i}\log(1-u_1) & R^\MHV_{6}\Big|_{\prod\zm_{a_i+1}}=\sum\limits_{\ell=n_z}^{\infty}\sum\limits_{m=\sum a_i}^\infty\sum\limits_{k=0}^\infty (-1)^m \frac{\eta^{\ell+m+k+1}}{4}\frac{2^{-k-m}}{\ell!\,k!\,m!}\CI\left[\xp{m}\xm{k}\Ep^\ell\right]_{\prod\zm_{a_i+1}}\nonumber\\
	&=\sum\limits_{\ell=n_z}^{\infty}\sum\limits_{m=\sum a_i}^\infty\sum\limits_{k=0}^\infty \frac{\eta^{\ell+m+k+1}}{4}\frac{(-1)^m 2^{-k-m}}{(\ell-n_z)!\,k!\,m!}(-2)^{n_z}\frac{\prod\limits_i\zm_{a_i+1}}{\prod\limits_{a_i}n_{a_i}!}\CI\left[\xp{m-\sum a_i}\xm{k}\Ep^{\ell-n_z}\right]_\zfree\nonumber\\
	&=\dots=(-1)^{w}\eta^{w+1}2^{2n_z-w}\frac{\prod\limits_i\zm_{a_i+1}}{\prod\limits_{a_i}n_{a_i}!}\frac{1}{2}\mathcal{X}x_1\sum\limits_{k=0}^\infty\left(\frac{\eta}{2}\right)^k\frac{x_0^k}{(k+w-n_z)!},\label{eq:res_zetas}
\end{align}
where we used \eqn{eq:zeta_rec} in the second step and in the last step we used
that the weight of our $\zm$-structure is given by $w=\sum\limits_i
a_i+n_z$.  To compare results, we make use of an alternative representation
of $\mathfrak{Z}(n,m)$ as given in \cite{Pennington:2012zj},
\begin{equation}
	\mathfrak{Z}(n,m)=\sum\limits_{\beta\in P(n,m)}\prod\limits_i\frac{(\zm_{2i+1})^{\beta_i}}{\beta_i!},
	\label{eq:altdef_z}
\end{equation}
where $P(n,m)$ runs over the set of $n$-tuples of non-negative integers with
sum $m$ such that the weight of the product of $\zm$'s is given by $n$.  This
means that the remainder function multiplying a given combination of $\zm$'s of
weight $w$ and consisting of $n_z$ $\zm$'s as obtained from
\eqn{eq:rlla} is given by
\begin{align}
	\frac{1}{2\pi i}\log(1-u_1)&R_6^{\mathrm{MHV}}\Big|_{\mathrm{LLA},\prod\limits_i (\zm_{2i+1})^{\beta_i}}\nnl
	&=\frac{1}{2}\mathcal{X}\sum\limits_{k=w+1}x_1 (-1)^w x_0^{k-w-1}\frac{2^{2n_z-k+1}}{(k-n_z-1)!}\left(\left.\mathfrak{Z}(w,n_z)\right|_{\prod(\zm_{2i+1})^{\beta_i}}\right)\eta^k\nnl
	&= (-1)^w \eta^{w+1}2^{2n_z-w}\left(\prod\limits_i\frac{(\zm_{2i+1})^{\beta_i}}{\beta_i!}\right)\frac{1}{2}\mathcal{X}x_1\sum\limits_{k=0}^\infty \left(\frac{\eta}{2}\right)^k\frac{x_0^k}{(k+w-n_z)!}.\nonumber
\end{align}
This exactly matches our result \eqn{eq:res_zetas}, which finishes the proof.


\section{Conclusions}
\label{sec:conclusion}

In this paper, we have presented an efficient algorithm for the calculation of
the six-point remainder function of planar $\CN=4$ super-Yang--Mills theory in
the multi-Regge regime.  So far, this calculation had to be carried out by
performing an inverse Fourier-Mellin transform by either direct calculation or
by making a suitable ansatz and matching series expansions, which becomes involved for a
large number of loops or a high logarithmic order.  The key improvement which our algorithm provides is
that it works directly in momentum space, without the need to make an ansatz or
to carry out any non-trivial integrals.  It is built on the observation made in
\rcite{Drummond:2015jea} that the full result is already encoded in a small
subset of the residues appearing in the Fourier-Mellin transform, working under
the assumption that the full result will be expressed through SVHPLs.  We use
this observation to derive recursion relations between different integrals, 
which allow to reduce any integral appearing in the expansion of the remainder
function to a small set of basis integrals.  These, in turn, can be trivially
evaluated in terms of Euler $Z$-sums.  The full result is then obtained from
the basis integrals by simply performing the natural actions on a SVHPL, namely
attaching an index and shuffling.  Using these results, we are able to proof
the formula for the MHV remainder function in LLA conjectured in
\rcite{Pennington:2012zj}.  Our algorithm can be combined with expressions for
the BFKL eigenvalue and impact factor in Fourier-Mellin space derived in
\rcite{Basso:2014pla}, which are conjectured to be exact, to generate data for
any number of loops at any logarithmic order in principle.  As specific
applications of this, we provide plots for the remainder function in N$^3$LLA
up to ten loops and for N$^4$LLA up to nine loops. The highest number of loops
for which we fully determine the remainder function is seven, but extending
this result to higher loops should be possible.  Based on this data, we also
write down conjectures for the remainder function in the collinear-Regge
regime.

Several directions for future research emerge from our results: It would be
interesting to study other contexts in which SVHPLs arise, most notably the
application to Mueller-Navelet jets \cite{DelDuca:2013lma}.  Furthermore, in
this paper we have focused on the six-point case.  However, similar dispersion
relations can be written down in the multi-Regge regime of higher-point
amplitudes, as well \cite{Bartels:2011ge,Bartels:2014jya}.  It would be
interesting to understand how our findings carry over to these situations.

\subsection*{Acknowledgments} 
We would like to thank Reinke Sven Isermann for collaboration in the early
stages of the project.  Furthermore, we are grateful to Claude Duhr, Georgios
Papathanasiou and Matteo Rosso for helpful discussions and comments on our
draft. The research of JB is supported in part by the SFB 647 “Raum-Zeit-
Materie. Analytische und Geometrische Strukturen” and the Marie Curie Network
GATIS (gatis.desy.eu) of the European Union’s Seventh Framework Programme
FP7-2007-2013 under grant agreement No. 317089.  The work of JB and MS is
partially supported by the Swiss National Science Foundation through the NCCR
SwissMap.


\section*{Appendix}
\appendix

\section{Differential equations for insertions of $\chi_\pm$}
\label{sec:deq_wdw}
In this appendix, we show schematically how to prove that the differential
equations \eqns{eq:wdw_int}{eq:wdw_ints} hold on the level of the
integral and not just on the level of the integrand. We begin by showing
\begin{equation}
	w^*\partial_{w^*}\CI\left[\xp{n_1}\xm{n_2}E_\psi^{n_3}\right]=-\CI\left[\xp{n_1-1}\xm{n_2}E_\psi^{n_3}\right].
	\label{eq:deq_wsdws}
\end{equation}
Extracting the left-hand side of \eqn{eq:deq_wsdws} is slightly subtle
since our prescription \eqn{eq:res_presc} is to set $w^*=1$ which would
na\"ively set the left-hand side to zero.  Instead we first have to calculate the
integral, then act with the differential operator and only then set $w^*=1$.
For this, it is enough to extract the contribution proportional to $\log w^*$,
which evaluates to $1$ when acted upon with the differential operator, as all
other contributions with $w^*$-dependence drop out once we set $w^*=1$.  We
find
\begin{align}
	w^*\partial_{w^*}\Ib&\left[\xp{n_1}\xm{n_2} E_\psi^{n_3}\right]\nnl
	&=\left.\left(w^*\partial_{w^*}\left( (-2i)\sum\limits_{n=1}^\infty(-1)^n\frac{(-1)^{n_1}}{n_1!}i^{n_2}D_\nu^{n_1}\left(\frac{w^{i\nu+\frac{n}{2}}(w^*)^{i\nu-\frac{n}{2}}}{\left(\nu+i\frac{n}{2}\right)^{n_2+1}}E_\psi^{n_3}\right)\right)\right)\right|_{w^*=1,\,\nu=-i\frac{n}{2}}\nnl
	&=(-1)\left( (-2i)\sum\limits_{n=1}^\infty(-1)^n\frac{(-1)^{n_1-1}}{(n_1-1)!}i^{n_2}D_\nu^{n_1-1}\left(\frac{w^{i\nu+\frac{n}{2}}}{\left(\nu+i\frac{n}{2}\right)^{n_2+1}}E_\psi^{n_3}\right)\right)\Bigg|_{\nu=-i\frac{n}{2}}\nnl
	&=-\Ib\left[\xp{n_1-1}\xm{n_2}E_\psi^{n_3}\right].
\end{align}
Similarly, we have
\begin{align}
	w^*\partial_{w^*}\Iz&\left[\xp{n_1}\xm{n_2}E_\psi^{n_3}\right]\nnl
	&=\left.\left(w^*\partial_{w^*} \frac{(-1)^{n_1+n_2}}{(n_1+n_2+1)!}D_\nu^{n_1+n_2+1}\left(w^{i\nu+\frac{n}{2}}(w^*)^{i\nu-\frac{n}{2}}E_\psi^{n_3}\right)\right)\right|_{w^*=1, \nu=0}\nnl
	&=\frac{(-1)^{n_1+n_2}}{(n_1+n_2)!}D_\nu^{n_1+n_2}\left(w^{i\nu+\frac{n}{2}}E_\psi^{n_3}\right)\Bigg|_{\nu=0}\nnl
	&=-\Iz\left[\xp{n_1-1}\xm{n_2}E_\psi^{n_3}\right],
\end{align}
which, using \eqn{eq:res_presc}, finishes the calculation. Showing
\begin{equation}
	w\partial_{w}\CI\left[\xp{n_1}\xm{n_2}E_\psi^{n_3}\right]=-\CI\left[\xp{n_1}\xm{n_2-1}E_\psi^{n_3}\right]
	\label{eq:deq_wdw}
\end{equation}
is slightly more difficult, since although we can set $w^*=1$ right away, we
have to take into account all contributions from $w$.  We begin with the
residue at $\nu=n=0$ and obtain
\begin{align}
	w\partial_w\Iz&\left[\xp{n_1}\xm{n_2}E_\psi^{n_3}\right]\nnl
	&=w\partial_w\left( (-i)\frac{(-1)^{n_1+n_2+1}}{(n_1+n_2+1)!}D_\nu^{n_1+n_2+1}\left(w^{i\nu}E_\psi^{n_2}\right)\right)\Bigg|_{\nu=0}\nnl
	&=\frac{(-1)^{n_1+n_2}}{(n_1+n_2+1)!}\sum\limits_{k=1}^{n_1+n_2+1}\binom{n_1+n_2+1}{k}k \log^{k-1}w\,D_\nu^{n_1+n_2+1-k}\left(E_\psi^{n_3}\right)\Bigg|_{\nu=0}\nnl
	&=\frac{(-1)^{n_1+n_2}}{(n_1+n_2)!}D_\nu^{n_1+n_2}\left(w^{i\nu}E_\psi^{n_2}\right)\Bigg|_{\nu=0}\nnl
	&=-\Iz\left[\xp{n_1}\xm{n_2-1}E_\psi^{n_3}\right].
	\label{eq:i0_wdw}
\end{align}
Moving on to the other residues we find
\begin{align}
	\Ib&\left[\xp{n_1}\xm{n_2-1}E_\psi^{n_3}\right]\nnl
	&=(-2i)\sum\limits_{n=1}^\infty (-1)^n\frac{(-1)^{n_1}}{n_1!}i^{n_2-1}D_\nu^{n_1}\left(\frac{w^{i\nu+\frac{n}{2}}}{\left(\nu-i\frac{n}{2}\right)^{n_2}}E_\psi^{n_3}\right)\Bigg|_{\nu=-i\frac{n}{2}}\nnl
	&=(-2i)\sum\limits_{n=1}^\infty (-1)^n\frac{(-1)^{n_1}}{n_1!}i^{n_2-1}D_\nu^{n_1}\left(\frac{w^{i\nu+\frac{n}{2}}}{\left(\nu-i\frac{n}{2}\right)^{n_2+1}}E_\psi^{n_3}\left(\nu-i\frac{n}{2}\right)\right)\Bigg|_{\nu=-i\frac{n}{2}}\nnl
	&=(-2i)\sum\limits_{n=1}^\infty (-1)^n\frac{(-1)^{n_1}}{n_1!}\left(-n i^{n_2}D_\nu^{n_1}\left(\frac{w^{i\nu+\frac{n}{2}}}{\left(\nu-i\frac{n}{2}\right)^{n_2}}E_\psi^{n_3}\right)+n_1 i^{n_2} D_\nu^{n_1-1}\left(\frac{w^{i\nu+\frac{n}{2}}}{\left(\nu-i\frac{n}{2}\right)^{n_2}}E_\psi^{n_3}\right)\right)\Bigg|_{\nu=-i\frac{n}{2}}\nnl
	&=-w\partial_w\Ib\left[\xp{n_1}\xm{n_2}E_\psi^{n_3}\right],
\end{align}
where the last step follows from expanding in the number of derivatives acting
on $w$ as in \eqn{eq:i0_wdw})  This establishes \eqn{eq:wdw_int}.

\bibliographystyle{nb}
\bibliography{\jobname}

\def\cprime{$'$}
\begin{thebibliography}{10}
\ifx\href\asklfhas\newcommand{\href}[2]{#2}\fi
\ifx\arxivref\asklfhas\newcommand{\arxivref}[2]{\href{http://arxiv.org/abs/#1}{#2}}\fi
\ifx\doiref\asklfhas\newcommand{\doiref}[2]{\href{http://dx.doi.org/#1}{#2}}\fi
\raggedright
\small
\parskip 0pt

\bibitem{Lipatov:1976zz}
L.~N.~Lipatov,
\textit{``{Reggeization of the Vector Meson and the Vacuum Singularity in
  Nonabelian Gauge Theories}''},
\textsf{Sov.~J.~Nucl.~Phys.~23,~338~(1976)}.

\bibitem{Fadin:1975cb}
V.~S.~Fadin, E.~A.~Kuraev and L.~N.~Lipatov,
\textit{``{On the Pomeranchuk Singularity in Asymptotically Free Theories}''},
\textsf{\doiref{10.1016/0370-2693(75)90524-9}{Phys.~Lett.~B60,~50~(1975)}}.

\bibitem{Kuraev:1976ge}
E.~A.~Kuraev, L.~N.~Lipatov and V.~S.~Fadin,
\textit{``{Multi - Reggeon Processes in the Yang-Mills Theory}''},
\textsf{Sov.~Phys.~JETP~44,~443~(1976)}.

\bibitem{Balitsky:1978ic}
I.~I.~Balitsky and L.~N.~Lipatov,
\textit{``{The Pomeranchuk Singularity in Quantum Chromodynamics}''},
\textsf{Sov.~J.~Nucl.~Phys.~28,~822~(1978)}.

\bibitem{Lipatov:1993yb}
L.~N.~Lipatov,
\textit{``{High-energy asymptotics of multicolor QCD and exactly solvable
  lattice models}''},
\texttt{\arxivref{hep-th/9311037}{hep-th/9311037}}.

\bibitem{Lipatov:1994xy}
L.~N.~Lipatov,
\textit{``{Asymptotic behavior of multicolor QCD at high energies in connection
  with exactly solvable spin models}''},
\textsf{JETP~Lett.~59,~596~(1994)}.

\bibitem{Faddeev:1994zg}
L.~D.~Faddeev and G.~P.~Korchemsky,
\textit{``{High-energy QCD as a completely integrable model}''},
\textsf{\doiref{10.1016/0370-2693(94)01363-H}{Phys.~Lett.~B342,~311~(1995)}},
\texttt{\arxivref{hep-th/9404173}{hep-th/9404173}}.

\bibitem{Bern:2005iz}
Z.~Bern, L.~J.~Dixon and V.~A.~Smirnov,
\textit{``{Iteration of planar amplitudes in maximally supersymmetric
  Yang-Mills theory at three loops and beyond}''},
\textsf{\doiref{10.1103/PhysRevD.72.085001}{Phys.~Rev.~D72,~085001~(2005)}},
\texttt{\arxivref{hep-th/0505205}{hep-th/0505205}}.

\bibitem{Bartels:2008sc}
J.~Bartels, L.~N.~Lipatov and A.~Sabio~Vera,
\textit{``{N=4 supersymmetric Yang Mills scattering amplitudes at high
  energies: The Regge cut contribution}''},
\textsf{\doiref{10.1140/epjc/s10052-009-1218-5}{Eur.~Phys.~J.~C65,~587~(2010)}},
\texttt{\arxivref{0807.0894}{arxiv:0807.0894}}.

\bibitem{Bartels:2008ce}
J.~Bartels, L.~N.~Lipatov and A.~Sabio~Vera,
\textit{``{BFKL Pomeron, Reggeized gluons and Bern-Dixon-Smirnov
  amplitudes}''},
\textsf{\doiref{10.1103/PhysRevD.80.045002}{Phys.~Rev.~D80,~045002~(2009)}},
\texttt{\arxivref{0802.2065}{arxiv:0802.2065}}.

\bibitem{DelDuca:2009au}
V.~Del~Duca, C.~Duhr and V.~A.~Smirnov,
\textit{``{An Analytic Result for the Two-Loop Hexagon Wilson Loop in N = 4
  SYM}''},
\textsf{\doiref{10.1007/JHEP03(2010)099}{JHEP~1003,~099~(2010)}},
\texttt{\arxivref{0911.5332}{arxiv:0911.5332}}.

\bibitem{DelDuca:2010zg}
V.~Del~Duca, C.~Duhr and V.~A.~Smirnov,
\textit{``{The Two-Loop Hexagon Wilson Loop in N = 4 SYM}''},
\textsf{\doiref{10.1007/JHEP05(2010)084}{JHEP~1005,~084~(2010)}},
\texttt{\arxivref{1003.1702}{arxiv:1003.1702}}.

\bibitem{Bartels:2010ej}
J.~Bartels, J.~Kotanski and V.~Schomerus,
\textit{``{Excited Hexagon Wilson Loops for Strongly Coupled N=4 SYM}''},
\textsf{\doiref{10.1007/JHEP01(2011)096}{JHEP~1101,~096~(2011)}},
\texttt{\arxivref{1009.3938}{arxiv:1009.3938}}.

\bibitem{Bartels:2012gq}
J.~Bartels, V.~Schomerus and M.~Sprenger,
\textit{``{Multi-Regge Limit of the n-Gluon Bubble Ansatz}''},
\textsf{\doiref{10.1007/JHEP11(2012)145}{JHEP~1211,~145~(2012)}},
\texttt{\arxivref{1207.4204}{arxiv:1207.4204}}.

\bibitem{Bartels:2013dja}
J.~Bartels, J.~Kotanski, V.~Schomerus and M.~Sprenger,
\textit{``{The Excited Hexagon Reloaded}''},
\texttt{\arxivref{1311.1512}{arxiv:1311.1512}}.

\bibitem{Bartels:2014ppa}
J.~Bartels, V.~Schomerus and M.~Sprenger,
\textit{``{Heptagon Amplitude in the Multi-Regge Regime}''},
\textsf{\doiref{10.1007/JHEP10(2014)067}{JHEP~1410,~67~(2014)}},
\texttt{\arxivref{1405.3658}{arxiv:1405.3658}}.

\bibitem{Bartels:2014mka}
J.~Bartels, V.~Schomerus and M.~Sprenger,
\textit{``{The Bethe roots of Regge cuts in strongly coupled $ \mathcal{N}=4 $
  SYM theory}''},
\textsf{\doiref{10.1007/JHEP07(2015)098}{JHEP~1507,~098~(2015)}},
\texttt{\arxivref{1411.2594}{arxiv:1411.2594}}.

\bibitem{Bartels:2011ge}
J.~Bartels, A.~Kormilitzin, L.~N.~Lipatov and A.~Prygarin,
\textit{``{BFKL approach and $2 \to 5$ maximally helicity violating amplitude
  in ${\cal N}=4$ super-Yang-Mills theory}''},
\textsf{\doiref{10.1103/PhysRevD.86.065026}{Phys.~Rev.~D86,~065026~(2012)}},
\texttt{\arxivref{1112.6366}{arxiv:1112.6366}}.

\bibitem{Prygarin:2011gd}
A.~Prygarin, M.~Spradlin, C.~Vergu and A.~Volovich,
\textit{``{All Two-Loop MHV Amplitudes in Multi-Regge Kinematics From Applied
  Symbology}''},
\textsf{\doiref{10.1103/PhysRevD.85.085019}{Phys.~Rev.~D85,~085019~(2012)}},
\texttt{\arxivref{1112.6365}{arxiv:1112.6365}}.

\bibitem{Bartels:2013jna}
J.~Bartels, A.~Kormilitzin and L.~Lipatov,
\textit{``{Analytic structure of the $n=7$ scattering amplitude in
  $\mathcal{N}=4$ SYM theory in the multi-Regge kinematics: Conformal Regge
  pole contribution}''},
\textsf{\doiref{10.1103/PhysRevD.89.065002}{Phys.~Rev.~D89,~065002~(2014)}},
\texttt{\arxivref{1311.2061}{arxiv:1311.2061}}.

\bibitem{Bartels:2014jya}
J.~Bartels, A.~Kormilitzin and L.~N.~Lipatov,
\textit{``{Analytic structure of the $n=7$ scattering amplitude in
  $\mathcal{N}=4$ theory in multi-Regge kinematics: Conformal Regge cut
  contribution}''},
\textsf{\doiref{10.1103/PhysRevD.91.045005}{Phys.~Rev.~D91,~045005~(2015)}},
\texttt{\arxivref{1411.2294}{arxiv:1411.2294}}.

\bibitem{Dixon:2012yy}
L.~J.~Dixon, C.~Duhr and J.~Pennington,
\textit{``{Single-valued harmonic polylogarithms and the multi-Regge limit}''},
\textsf{\doiref{10.1007/JHEP10(2012)074}{JHEP~1210,~074~(2012)}},
\texttt{\arxivref{1207.0186}{arxiv:1207.0186}}.

\bibitem{Pennington:2012zj}
J.~Pennington,
\textit{``{The six-point remainder function to all loop orders in the
  multi-Regge limit}''},
\textsf{\doiref{10.1007/JHEP01(2013)059}{JHEP~1301,~059~(2013)}},
\texttt{\arxivref{1209.5357}{arxiv:1209.5357}}.

\bibitem{DelDuca:2013lma}
V.~Del~Duca, L.~J.~Dixon, C.~Duhr and J.~Pennington,
\textit{``{The BFKL equation, Mueller-Navelet jets and single-valued harmonic
  polylogarithms}''},
\textsf{\doiref{10.1007/JHEP02(2014)086}{JHEP~1402,~086~(2014)}},
\texttt{\arxivref{1309.6647}{arxiv:1309.6647}}.

\bibitem{Lipatov:2010ad}
L.~N.~Lipatov and A.~Prygarin,
\textit{``{BFKL approach and six-particle MHV amplitude in N=4 super
  Yang-Mills}''},
\textsf{\doiref{10.1103/PhysRevD.83.125001}{Phys.~Rev.~D83,~125001~(2011)}},
\texttt{\arxivref{1011.2673}{arxiv:1011.2673}}.

\bibitem{Fadin:2011we}
V.~S.~Fadin and L.~N.~Lipatov,
\textit{``{BFKL equation for the adjoint representation of the gauge group in
  the next-to-leading approximation at N=4 SUSY}''},
\textsf{\doiref{10.1016/j.physletb.2011.11.048}{Phys.~Lett.~B706,~470~(2012)}},
\texttt{\arxivref{1111.0782}{arxiv:1111.0782}}.

\bibitem{Lipatov:2012gk}
L.~Lipatov, A.~Prygarin and H.~J.~Schnitzer,
\textit{``{The Multi-Regge limit of NMHV Amplitudes in N=4 SYM Theory}''},
\textsf{\doiref{10.1007/JHEP01(2013)068}{JHEP~1301,~068~(2013)}},
\texttt{\arxivref{1205.0186}{arxiv:1205.0186}}.

\bibitem{Dixon:2014voa}
L.~J.~Dixon, J.~M.~Drummond, C.~Duhr and J.~Pennington,
\textit{``{The four-loop remainder function and multi-Regge behavior at NNLLA
  in planar N = 4 super-Yang-Mills theory}''},
\textsf{\doiref{10.1007/JHEP06(2014)116}{JHEP~1406,~116~(2014)}},
\texttt{\arxivref{1402.3300}{arxiv:1402.3300}}.

\bibitem{Basso:2014pla}
B.~Basso, S.~Caron-Huot and A.~Sever,
\textit{``{Adjoint BFKL at finite coupling: a short-cut from the collinear
  limit}''},
\textsf{\doiref{10.1007/JHEP01(2015)027}{JHEP~1501,~027~(2015)}},
\texttt{\arxivref{1407.3766}{arxiv:1407.3766}}.

\bibitem{Basso:2013vsa}
B.~Basso, A.~Sever and P.~Vieira,
\textit{``{Spacetime and Flux Tube S-Matrices at Finite Coupling for N=4
  Supersymmetric Yang-Mills Theory}''},
\textsf{\doiref{10.1103/PhysRevLett.111.091602}{Phys.~Rev.~Lett.~111,~091602~(2013)}},
\texttt{\arxivref{1303.1396}{arxiv:1303.1396}}.

\bibitem{Basso:2013aha}
B.~Basso, A.~Sever and P.~Vieira,
\textit{``{Space-time S-matrix and Flux tube S-matrix II. Extracting and
  Matching Data}''},
\textsf{\doiref{10.1007/JHEP01(2014)008}{JHEP~1401,~008~(2014)}},
\texttt{\arxivref{1306.2058}{arxiv:1306.2058}}.

\bibitem{Basso:2014koa}
B.~Basso, A.~Sever and P.~Vieira,
\textit{``{Space-time S-matrix and Flux-tube S-matrix III. The two-particle
  contributions}''},
\textsf{\doiref{10.1007/JHEP08(2014)085}{JHEP~1408,~085~(2014)}},
\texttt{\arxivref{1402.3307}{arxiv:1402.3307}}.

\bibitem{Basso:2014nra}
B.~Basso, A.~Sever and P.~Vieira,
\textit{``{Space-time S-matrix and Flux-tube S-matrix IV. Gluons and
  Fusion}''},
\textsf{\doiref{10.1007/JHEP09(2014)149}{JHEP~1409,~149~(2014)}},
\texttt{\arxivref{1407.1736}{arxiv:1407.1736}}.

\bibitem{Brown2004527}
F.~C.~Brown,
\textit{``Polylogarithmes multiples uniformes en une variable''},
\textsf{\doiref{http://dx.doi.org/10.1016/j.crma.2004.02.001}{Comptes~Rendus~Mathematique~338,~527
  ~(2004)}},
\href{http://www.sciencedirect.com/science/article/pii/S1631073X04000780}{\texttt{http://www.sciencedirect.com/science/article/pii/S1631073X04000780}}.

\bibitem{Drummond:2015jea}
J.~M.~Drummond and G.~Papathanasiou,
\textit{``{Hexagon OPE Resummation and Multi-Regge Kinematics}''},
\texttt{\arxivref{1507.08982}{arxiv:1507.08982}}.

\bibitem{Drummond:2013nda}
J.~Drummond, C.~Duhr, B.~Eden, P.~Heslop, J.~Pennington and V.~A.~Smirnov,
\textit{``{Leading singularities and off-shell conformal integrals}''},
\textsf{\doiref{10.1007/JHEP08(2013)133}{JHEP~1308,~133~(2013)}},
\texttt{\arxivref{1303.6909}{arxiv:1303.6909}}.

\bibitem{Dixon:2014iba}
L.~J.~Dixon and M.~von~Hippel,
\textit{``{Bootstrapping an NMHV amplitude through three loops}''},
\textsf{\doiref{10.1007/JHEP10(2014)065}{JHEP~1410,~065~(2014)}},
\texttt{\arxivref{1408.1505}{arxiv:1408.1505}}.

\bibitem{Beisert:2006ez}
N.~Beisert, B.~Eden and M.~Staudacher,
\textit{``{Transcendentality and Crossing}''},
\textsf{\doiref{10.1088/1742-5468/2007/01/P01021}{J.~Stat.~Mech.~0701,~P01021~(2007)}},
\texttt{\arxivref{hep-th/0610251}{hep-th/0610251}}.

\bibitem{Papathanasiou:2014yva}
G.~Papathanasiou,
\textit{``{Evaluating the six-point remainder function near the collinear
  limit}''},
\textsf{\doiref{10.1142/S0217751X14501541}{Int.~J.~Mod.~Phys.~A29,~G.~Papathanasiou~(2014)}},
\texttt{\arxivref{1406.1123}{arxiv:1406.1123}},
in: \textit{``Proceedings, 49th Rencontres de Moriond on QCD and High Energy
  Interactions''},
ed.: {E. Aug\'{e}, J. Dumarchez, J. T. T. V\^{a}n},
ARISF (2014).

\bibitem{Moch:2001zr}
S.~Moch, P.~Uwer and S.~Weinzierl,
\textit{``{Nested sums, expansion of transcendental functions and multiscale
  multiloop integrals}''},
\textsf{\doiref{10.1063/1.1471366}{J.~Math.~Phys.~43,~3363~(2002)}},
\texttt{\arxivref{hep-ph/0110083}{hep-ph/0110083}}.

\bibitem{Bauer:2000cp}
C.~W.~Bauer, A.~Frink and R.~Kreckel,
\textit{``{Introduction to the GiNaC framework for symbolic computation within
  the C++ programming language}''},
\textsf{J.~Symb.~Comput.~33,~1~(2000)},
\texttt{\arxivref{cs/0004015}{cs/0004015}}.

\bibitem{Bartels:2011xy}
J.~Bartels, L.~N.~Lipatov and A.~Prygarin,
\textit{``{Collinear and Regge behavior of 2 $\rightarrow$ 4 MHV amplitude in N
  = 4 super Yang-Mills theory}''},
\texttt{\arxivref{1104.4709}{arxiv:1104.4709}}.

\end{thebibliography}

\end{document}